\documentclass[prd,
 reprint,
 amsmath,amssymb,
 aps,nofootinbib
]{revtex4-2}
\usepackage{bm}% bold math
\PassOptionsToPackage{linktocpage}{hyperref}
\usepackage[hyperindex,breaklinks,hidelinks]{hyperref}
\usepackage{enumitem}
\usepackage{slashed}
\usepackage[dvipsnames]{xcolor}
\newcommand{\Lag}{\mathcal{L}}

\usepackage{subcaption}

\usepackage{microtype}

\usepackage[compat=1.1.0]{tikz-feynman}

\newcommand{\Ll}{\mathcal{L}} % Lagrangian

\usepackage{tikz}
\tikzset{every picture/.style={line width=0.75pt}} %set default line width to 0.75pt        
\usetikzlibrary{shapes.geometric}
\usepackage{orcidlink}
\usepackage{comment}
\usepackage{array}
\usepackage{mathtools}

\usepackage{etoolbox}

\begin{document}

\title{Cosmic strings and domain walls of the QCD quark condensate \\
with and without a hidden axion.}
%\title{Cosmic strings and domain walls of $\eta'$-axion} 

\author{Gia Dvali} 
\email{gia.dvali@mpp.mpg.de}
\author{Lucy Komisel\,\orcidlink{0009-0008-1578-588X}} 
\email{lucy.komisel@mpp.mpg.de}
\author{Anja Stuhlfauth\,\orcidlink{0009-0005-0920-379X}} 
\email{anja.stuhlfauth@mpp.mpg.de}
\affiliation{Arnold Sommerfeld Center, Ludwig-Maximilians-Universit\"at, Theresienstra{\ss}e 37, 80333 M\"unchen, Germany}
\affiliation{Max-Planck-Institut f\"ur Physik, Boltzmannstr. 8, 85748 Garching, Germany}

\date{\today}

\begin{abstract} 

The chiral quark condensate of QCD, which spontaneously breaks the anomalous axial symmetry, gives rise to axionic type global string-wall systems. If a Peccei-Quinn type axion exists in the theory, the axionic strings are in general accompanied by winding of the QCD quark condensate. Depending on the axion model the winding can proceed either in the $\eta'$ or in the pion direction. This determines the structure of fermionic zero modes and the anomaly inflow which has important astrophysical consequences. We point out that $\eta'$ and pion string-wall systems exist in pure QCD, independently of the hidden axion. 
Strikingly, even if a hidden axion exists, the early cosmology can be entirely dominated by string-wall systems formed by the QCD quark condensate.
We also discuss their role in the QCD phase transition and in heavy-ion physics.

% VERSION2:
%  $\eta'$-meson of QCD is  a full-fledged axion, albeit of 
%  poor quality due to non-zero quark masses.  
%    We discuss, domain walls and global cosmic strings composed of $\eta'$. These can exist independently 
%  of the existence of a good-quality axion. If a good quality axion exists, $\eta'$ becomes a intrinsic part of the axionic string-wall network.  In particular, in the lowest energy state,  
%  $\eta'$ winds around the Peccei-Quinn axionic string 
%  implying that the axion string is surrounded by 
%  the QCD core with vanishing quark condensate. 
%   The winding of $\eta'$ localizes the zero modes of 
%   the light quarks that cancel the anomaly. 

\end{abstract}

\maketitle

\section{Introduction}

The chiral symmetry of quarks is spontaneously broken by the chiral quark condensate. At the same time, this chiral symmetry is explicitly broken by quark masses as well as by the chiral anomaly. Due to the explicit breaking, the would-be massless Nambu-Goldstone bosons, such as the pions and the $\eta'$-meson, gain non-zero masses. Since no spontaneously broken discrete subgroup survives this explicit breaking, the vacuum of the chiral symmetry is topologically trivial without truly stable topological defects. However, from the point of view of various cosmological and phenomenological implications the unstable defects are no less interesting.
 
In this work, we point out that, despite the above limitations, there exist non-trivial winding configurations corresponding to $2\pi$-domain walls bounded by strings. The windings take place in two major flavor directions corresponding to the pseudo-Nambu-Goldstone bosons $\eta'$ or $\pi^0$. However, winding in other directions such as the $\eta$-meson is also possible.    
  
If a hidden Peccei-Quinn QCD axion exists in nature, the axionic strings and walls are in general accompanied by windings of the phases of the QCD quark condensate. This has important implications for the zero mode content of axionic strings and, correspondingly, for 
their superconductivity properties \cite{Witten:1984eb} as well as the dynamics of the anomaly inflow \cite{Callan:1984sa}. 
 
However, more importantly, the QCD $2\pi$-string-wall systems can exist regardless of the hidden axion. As we shall explain in detail, a simple way to understand this is that the $\eta'$-meson of QCD represents an axion, albeit of poor quality, which directly probes the physics of the QCD $\theta$-vacuum. In particular, in the limit of a vanishing quark mass, the $\eta'$-meson fully cancels the vacuum $\theta$-angle \cite{Dvali:2005an}.   
The non-zero quark mass offsets this cancellation but does preserve the axion-like nature of the $\eta'$-meson. Correspondingly, the $\eta'$-meson mirrors the structures usually discussed only in the context of a hidden axion, such as domain walls bounded by cosmic strings.   
 
Since the vacuum of QCD is $2\pi$-periodic in the phases of the quark condensate, the phases change across the wall by multiples of $2\pi$. We shall refer to such domain walls as $2\pi$-walls. A planar infinite wall is classically stable. Although via a quantum tunneling process a hole can be punched through it, the probability is highly suppressed. Of course, in reality, infinite walls never form. Instead, they materialize during phase transitions in the form of closed bubbles or as walls bounded by cosmic strings. Such structures collapse and radiate energy in various modes, including electromagnetic radiation as well as gravitational waves. 
    
String-wall systems formed by the QCD condensate can have particularly important cosmological implications in scenarios with an early epoch QCD phase transition \cite{Dvali:1995ce}.  
 
From the low energy perspective of the theory of mesons, the QCD condensate string-wall systems represent solitonic states. We discuss their potential manifestations in heavy-ion experiments.

 \section{Strong-CP puzzle} 
 
In the standard discussion (which ignores gravity) the strong-$CP$ puzzle is formulated as the following naturalness problem.  
QCD has a continuum of vacua conventionally labeled by the $CP$-violating vacuum angle $\theta$~\cite{Callan:1976je,Jackiw:1976pf}.
 
These vacua belong to different superselection sectors. 
The choice of the vacuum is accounted for by the following term in the Lagrangian, 
\begin{equation} \label{thetaterm}
     \frac{g^2}{32 \pi^2} \, \bar{\theta} \, G_{\mu\nu} \tilde{G}^{\mu\nu}  \,,
\end{equation}
where $G_{\mu\nu}$ is the gluon field strength and $ \tilde{G}^{\mu\nu} \equiv \epsilon^{\mu\nu\alpha\beta} G_{\alpha\beta}$ is its dual. In the rest of this paper, we will drop the constants in front, including the strong coupling constant $g$.

In a theory with massive quarks, the physically measurable parameter is the quantity $\bar{\theta} \equiv \theta  + {\rm arg(det\ M_q}) $,
where ${\rm arg(det\ M_q)}$ is the phase of the determinant of the quark mass matrix. 

In the quantum theory, the $\bar{\theta}$-term induces an electric dipole moment of the neutron (nEDM) \cite{Baluni:1978rf,Crewther:1979pi}, denoted as $d_n$. 
The comparison of the resulting theoretical value with the current experimental limit,  $|d_n| <  2.9 \times 10^{-26} e$cm \cite{Baker:2006ts, Pendlebury:2015lrz, Graner:2016ses}, gives the bound
\begin{equation} \label{BBartheta}
    |\bar{\theta}|  \lesssim 10^{-9}\,. 
\end{equation} 
Notice that there exists an additional contribution to the nEDM coming from the breaking of $CP$-symmetry by the weak interaction \cite{Ellis:1976fn, Shabalin:1978rs, Ellis:1978hq}.
However, this correction is too small to affect the current bound (\ref{BBartheta}).
  
Thus, observations indicate that we live in a sector with a minuscule or zero $\bar{\theta}$. 
This is puzzling.

\section{Peccei-Quinn solution} 

Let us consider the Peccei-Quinn solution \cite{Peccei:1977hh,*Peccei:1977ur} which is based on removing  $\bar{\theta}$ by an anomalous chiral
symmetry $U(1)_{PQ}$. For simplicity, let us consider a single quark  flavor, $\Psi$, that transforms under the symmetry as 
\begin{equation} \label{eq:Upsi}
\Psi \rightarrow {\rm e}^{i\frac{1}{2}\alpha \gamma_5} \Psi\,.
\end{equation} 
Due to the chiral anomaly, the current exhibits an anomalous divergence: 
\begin{equation} %\label{E1}
    \partial^{\mu} (\bar\Psi  \gamma_{\mu} \gamma_5 \Psi)\
    \propto G \tilde{G}.
\end{equation} 
Correspondingly, the Lagrangian shifts as 
\begin{equation} %\label{E2}
    \delta \Lag \, \propto \,  \alpha \,  G \tilde{G}.
\end{equation} 
Thus, the $\bar{\theta}$-term can be removed by the chiral
$U(1)_{PQ}$ transformation, 
\begin{equation} %\label{E2}
    \bar{\theta}  \rightarrow  \bar{\theta} - \alpha.
\end{equation} 
This indicates that $\bar{\theta}$ is unphysical. 

The fact that, in the presence of an anomalous chiral symmetry, 
the $\bar{\theta}$-term is unphysical can also be understood in the language of instanton zero modes. According to the index theorem, each spin-$1/2$ quark deposits a chiral zero mode in the instanton background.  
These modes kill the instanton effect. 

However, in nature there exist no massless quarks. 
Therefore, the $U(1)_{PQ}$-symmetry must be spontaneously 
broken by the vacuum expectation value (VEV) of  a complex scalar field 
$\Phi(x)= \rho(x) {\rm e}^{i \theta_{\phi}(x)}$,   
which under the $U(1)_{PQ}$ transforms as  
\begin{equation} \label{eq:uphi}
    \Phi \rightarrow {\rm e}^{i\alpha} \Phi\,. 
\end{equation}
 
The modulus of the scalar gets a VEV, $\langle \rho \rangle  = f_{\phi}$, from the Mexican-hat (Goldstone) potential  
\begin{equation} %\label{potPQ} 
    V(\Phi) = \lambda^2 (\Phi^*\Phi -f_{\phi}^2)^2 \,,
\end{equation} 
where $\lambda$ is a coupling constant. 
This gives a mass, $M_{\Psi} = g_{\Psi} f_{\phi}$, to the fermion through a Yukawa coupling in the fermion Lagrangian 
\begin{equation} \label{eq:LPQ}
    \Lag_{\Psi} \, = \, i \bar{\Psi}_{L,R} \gamma^{\mu} D_{\mu} \Psi_{L,R}  - g_\Psi \Phi^*  \, \bar{\Psi}_L  \Psi_R \,  + {\rm h.c}.,
\end{equation}
where we have written the Dirac spinors 
$\Psi = \Psi_{L} + \Psi_{R}$
in terms of left-right components,   $g_\Psi$ is the Yukawa coupling constant and 
$D_{\mu}$ is the covariant derivative with respect to the gauge symmetry. 

The Nambu-Goldstone phase, $\theta_{\phi}(x)$, is the axion \cite{Weinberg:1977ma, *Wilczek:1977pj}, which in terms of a canonically normalized field $\phi(x)$ can be written as $\theta_{\phi}(x) \equiv \frac{\phi(x)}{\sqrt{2} f_{\phi}}$.  
   
Taking the anomaly into account, the effective theory of the axion is\footnote{Notice that, since in this simplified discussion there are no other quarks, $\bar{\theta} = \theta$.}:
\begin{equation} \label{EcosA}
    \Lag_{\rm axion} = \, f_\phi^2(\partial_{\mu} \theta_{\phi}(x))^2 \, -  \left (\theta_{\phi}(x) - \bar{\theta} \right ) G\tilde{G}.
\end{equation}
Thus, the effective vacuum angle, $\theta_{\rm eff}(x)  :=  \theta_{\phi}(x) - \bar{\theta}$, becomes dynamical.    
The axion relaxes dynamically to the global minimum 
which, according to the Vafa-Witten theorem~\cite{Vafa:1984xg}, is at $\theta_{\rm eff}  = 0$.

This can be seen by an explicit computation of the energy dependence on $\theta_{\rm eff}(x)$.
Indeed, non-perturbative effects (instantons) generate a potential for the axion.  In the dilute instanton gas approximation,  
\begin{equation} %\label{PQNM2} 
    V(\theta_\phi) =  - \Lambda^4\,  \cos\left (\theta_{\phi}(x) - \bar{\theta}\right ), 
\end{equation} 
which has a minimum at $\theta_{\rm eff} =0$.

\section{$\eta'$-meson as poor-quality axion}  

Interestingly, QCD already contains an axion, albeit of poor quality. This axion is the $\eta'$-meson.         
Consider QCD with $N_f$ flavors of massless (or light) quarks $\psi^{i}$, where $i=1,2, ...N_f$ is the flavor index. 
Each flavor represents a Dirac fermion, 
$\psi^{i} = \psi^{i}_L + \psi^{i}_R$, with both left-handed and right-handed chiralities transforming as fundamental representations of the color group.

The term "light" refers to the quarks with masses below the QCD scale $\Lambda$.  
In the real world, $N_f = 3$, since there are three light quarks, $u,d,s$.
However, let us keep the discussion more general and assume for now that these $N_f$ light quark flavors are massless.   

In the case of massless quarks, QCD exhibits a global $U(N_f)_L \times U(N_f)_R$ symmetry.
In reality, this symmetry is only approximate because it is explicitly broken by quark masses.
One particularly interesting subgroup of this global symmetry group is the axial $U(1)_A$-symmetry, under which the quarks transform as
\begin{equation} \label{eq:axialtransformation}
    \psi_i \rightarrow {\rm e}^{-i\frac{1}{2}\alpha \gamma_5} \psi_i.
\end{equation} 
  
The $U(1)_A$-symmetry is spontaneously broken by the quark condensate $\langle \bar{\psi}_i \psi_j\rangle = \Lambda^3\delta_{ij}$. 
However, the corresponding massless (or light) Nambu-Goldstone boson,  residing in  $\bar{\psi}_i \gamma_5 \psi_i$,    
was nowhere  to be found. This is the famous axial $U(1)_A$-problem. 
       
This problem was solved by 't Hooft \cite{tHooft:1976rip,*tHooft:1976snw}, who understood that the would-be Goldstone boson was getting mass from the $U(1)_A$-anomaly through instantons. 
The resulting potential for the phase degree of freedom, $\theta_{\eta}(x)$, in the dilute instanton gas approximation is: 
\begin{equation} \label{eq:etapotential} 
    V(\eta') =  - \Lambda^4\,  \cos\left (N_f \,  \theta_{\eta} - \theta\right ). 
\end{equation}   
The phase of the condensate, $ \theta_{\eta}(x) = \frac{\eta'(x)}{\sqrt 2 f_{\eta}}$ is related to the canonically normalized $\eta'$ field through $f_{\eta}$, the decay constant of $\eta'$. The potential \eqref{eq:etapotential} gives a mass to $\eta'$, $m_{\eta} \sim \frac{\Lambda^2}{f_{\eta}}$.
In the real world, with $3$ colors and $3$ light quark flavors,  $f_{\eta} \sim \Lambda$.   
  
Interestingly, 't Hooft did not point out that simultaneously with generating a mass for $\eta'$, the potential gives $\theta_{\rm eff} =0$ and, therefore, solves the strong-$CP$ problem. 
 
In other words, in QCD with massless quarks, $\eta'$ is a full-fledged axion \cite{Dvali:2005an,Dvali:2005ws,Dvali:2013cpa}.
Unfortunately (or fortunately?), in the real world quarks are massive \cite{FlavourLatticeAveragingGroupFLAG:2024oxs}, which explicitly breaks $U(1)_A$. Therefore, $\eta'$ cannot enforce $\theta_{\rm eff} = 0$. This can be seen from the effective theory obtained after inclusion of the quark masses. For simplicity, we consider a single light quark with the following mass term, 
\begin{equation} %\label{PQNM} 
    m_{\psi} \bar{\psi}_L{\psi}_R + {\rm h.c.}
\end{equation}   
After the inclusion of the mass term, the effective potential of the phase degree of freedom becomes
\begin{equation}% \label{PQNM}
    V(\eta') =  - \Lambda^4\,  \cos\left ( \theta_{\eta} - \bar{\theta}\right )  \, -  \,\Lambda_{m}^4\,  \cos\left ( \theta_{\eta}\right ),
\end{equation}   
where $\Lambda_{m}^4 \simeq m_{\psi} \Lambda^3$. 
We do not take any other neutral mesons into account, as we are only working with one light flavor.
Here, we included the phase of the mass, $\theta_m$, into $\bar{\theta} = \theta + \theta_{m}$.
The minimum of the potential is now at 
\begin{equation}
    \theta_{\eta} = \arctan \left (\tan(\bar{\theta}) \bigg(1 + \frac{\Lambda_m^4}{\Lambda^4\cos(\bar{\theta})} \bigg)^{-1} \right ).
\end{equation}
Assuming that $\bar{\theta}$ is small\footnote{This assumption is made for illustrative purposes, there is no physical reason why $\bar\theta$ would have to be small. The result that $\eta'$ turns $\theta_{\rm eff}=0$ for massless quarks holds for any value of $\bar\theta$.}
such that we can use $\tan \bar{\theta} \approx \bar{\theta}$, the effective vacuum angle $\theta_{\mathrm{eff}} = \theta_\eta -\bar\theta$ is 
\begin{equation} %\label{PQNM} 
    \theta_{\rm eff} \simeq -\frac{|m_{\psi}|}{\Lambda} \bar{\theta}.
\end{equation}   
 All observable $CP$-violating effects, such as nEDM, are controlled by $\theta_{\rm eff}$.
Thus, in the case of massless quarks, $\eta'$ turns $\theta_{\rm eff}=0$ and solves the strong-CP puzzle.
However, since in the real world quarks are massive with $m_\psi/\Lambda\sim 10^{-2}$, $\eta'$~cannot provide an explanation for the observed smallness of $\theta_{\rm eff}$.
In other words, for $m_{\psi} \neq 0$, $\eta'$ is a poor-quality axion. 
  
Nevertheless, the lessons we learn from above are extraordinarily important: 
   
{\it{1)}} The mass of $\eta'$ represents an experimental proof of the existence of $\theta$-vacua. 
    
{\it{2})} $\eta'$ illustrates the reality of axion dynamics:  
$\eta'$ would be an exact quality axion, if at least one quark would be massless \cite{Dvali:2005an}. 
    
{\it{3)}} Since quarks are massive, there must exist a good-quality axion.

\section{Coupled system} 

In order to understand the mixing between the $\eta'$-meson and the PQ axion field, we consider a simplified model that includes a single light quark $\psi$ and a single heavy quark $\Psi$. The latter gets its mass from the PQ field $\Phi$. In the low energy EFT this model gives rise to an axion and the $\eta '$-meson. 
The simplest way to understand their 
coupling is through the 't Hooft determinant. 
       
\subsection{Massless quark case} 

We first consider the case where the explicit mass term of the light quark, $m_\psi$, vanishes.
The tree-level UV Lagrangian has the form
\begin{align} \label{LUV}
%\label{LPQ2}
    \Lag_{\rm UV} =& \, \partial_{\mu} \Phi^* \partial^{\mu} \Phi - \lambda^2 (\Phi^*\Phi -f_{\Phi}^2)^2 + \\
    \nonumber+& \, i \bar{\Psi}_{L,R} \gamma^{\mu} D_{\mu} \Psi_{L,R}  - g_\Psi \Phi^* \bar{\Psi}_L  \Psi_R + {\rm h.c.}\\
    \nonumber+&   i \bar{\psi}_{L,R} \gamma^{\mu} D_{\mu} \psi_{L,R}  - m_{\psi} \bar{\psi}_L  \psi_R + {\rm h.c.} - \\
    -& G_{\mu\nu}G^{\mu\nu} + \bar{\theta} \,  G_{\mu\nu}\tilde{G}^{\mu\nu},\nonumber 
\end{align} 
(irrelevant numerical factors are dropped for simplicity). 
We start by analyzing the theory for $m_{\psi} = 0$ and switch on this explicit mass term later. 
 
For $m_{\psi} = 0$, the theory contains two interesting global $U(1)$-symmetries: the axial symmetry in \eqref{eq:axialtransformation}
and the PQ symmetry in \eqref{eq:Upsi} and \eqref{eq:uphi}. We can combine them to get an anomalous axial symmetry, which we shall denote by $U(1)_A$,  
\begin{equation} %\label{U1A}
    \Phi \rightarrow {\rm e}^{i\alpha} \Phi, \  \Psi \rightarrow {\rm e}^{i\frac{1}{2}\alpha \gamma_5} \Psi,
  \ \psi \rightarrow {\rm e}^{i\frac{1}{2}\alpha \gamma_5} \psi,
\end{equation} 
and the anomaly-free $U(1)_V$ symmetry,
\begin{equation} %\label{U1V}
   \Phi \rightarrow {\rm e}^{i\alpha} \Phi\,,\ \Psi \rightarrow {\rm e}^{i\frac{1}{2}\alpha \gamma_5} \Psi\,,
    \ \psi \rightarrow {\rm e}^{-i\frac{1}{2}\alpha \gamma_5} \psi\,.
\end{equation} 
Both symmetries are spontaneously broken by two sources: 
The VEV of the field $\langle |\Phi| \rangle = f_{\phi}$ and the QCD condensate of the light quark $\langle \bar{\psi}{\psi} \rangle = \Lambda^3$. 
 
As previously, the associated phase degrees of freedom we shall denote by $\theta_{\phi}$  and $ \theta_{\eta}$ respectively.
They can be expressed through corresponding canonically-normalized Nambu-Goldstone bosons and their decay constants as 
\begin{equation} %\label{Golds}
    \theta_{\phi} = \frac{\phi}{\sqrt{2}f_{\phi}}\,, ~~~\,  \theta_{\eta} = \frac{\eta'}{\sqrt{2}f_{\eta}} \,. 
\end{equation} 
 
However, the anomalous symmetry $U(1)_A$ is explicitly broken by instantons.    
After taking into account the $U(1)_A$-violating instanton effect the effective Lagrangian takes the form
\begin{align} \label{eq:LGoldMzero}
    \Lag_{\rm eff} &=  f_{\phi}^2 (\partial_{\mu} \theta_{\phi} \partial^{\mu} \theta_{\phi}) + f_{\eta}^2  (\partial_{\mu} \theta_{\eta} \partial^{\mu} \theta_{\eta}) \\
  \nonumber &+  \Lambda^4  \cos\left (\theta_{\phi} +  \theta_{\eta}  -  \bar{\theta} \right ) .
\end{align} 
The cosine potential comes from the dilute instanton gas approximation, which suffices for illustrating our point.  
The exact form of the potential is unimportant. The only essential information is that the potential is a periodic function with minimum at $\bar{\theta}_{\rm eff} =0$.
  
The phase degree of freedom that cancels $\bar{\theta}$  is $\theta_{\phi} + \theta_{\eta}$ which represents the Goldstone phase of the spontaneously broken axial $U(1)_A$-symmetry. 
In  terms of the canonically normalized field, which we shall denote by $a_{\eta}$, it can be written as: 
\begin{equation} 
    \theta_{\phi} +  \theta_{\eta} =   \frac{a_{\eta}}{\sqrt{2} \tilde{f}}
  \end{equation} 
where
\begin{equation}
    \tilde{f} \equiv  \frac{f_{\phi}f_{\eta}}{f}\quad {\rm and}\quad f=\sqrt{f_{\phi}^2 + f_{\eta}^2}.
\end{equation}
The degree of freedom $a_\eta$ is a combination of the $\eta'$ and $\phi$,
\begin{equation}
    a_{\eta} = \eta' \cos(\xi)  + \phi \sin(\xi)
\end{equation}
with mixing angle $\xi$ given by $\sin(\xi) = f_{\eta} /f$.
Since $f_{\phi} \gg f_{\eta}$, it is clear that the mixing angle is tiny, $\xi \simeq \frac{f_{\eta}}{f_{\phi}} \ll 1$. 
Correspondingly, the decay constant is $\tilde{f} \simeq f_{\eta}$ and the true axion,
\begin{equation} 
    \label{eq:aeta}
    a_{\eta}  \simeq \eta'   + \xi \phi  \,, 
\end{equation} 
is mostly made out of $\eta'$ with a negligible ${\mathcal O}(\xi)$ admixture from $\phi$. The mass of this particle is generated from the instanton term and is 
\begin{equation} \label{eq:masseta}
    m_{\eta} \simeq \frac{\Lambda^2}{f_{\eta}}  \,.
\end{equation}  
The orthogonal degree of freedom, 
\begin{equation} 
    \label{eq:aphi}
    a_{\phi}  \equiv  \phi  \cos(\xi) - \eta' \sin(\xi) \simeq \phi  - \xi \eta'\,,   
\end{equation} 
which is predominantly made out of $\phi$, remains massless.
This field corresponds to a Goldstone boson of the spontaneously broken $U(1)_V$ symmetry.
Since the $U(1)_V$ is anomaly free and thus not explicitly broken by instantons, its Goldstone boson remains massless.

Thus, when the theory contains a massless quark, the axion is the $\eta'$-meson \cite{Dvali:2005an}, 
whereas the phase of the PQ field, $\phi$, remains massless and plays no role in canceling 
$\bar{\theta}$.

\subsection{Massive quark case} 
  
In order to account for the effect of a non-zero mass of the light quark, we now deform the theory by putting $m_{\psi} \neq 0$.
As a result, the theory possesses no spontaneously broken global symmetry spared of explicit breaking. 
Correspondingly, both Goldstone bosons become massive. 
  
To analyze the vacuum structure, we study the effective Lagrangian of the phases which now takes the form,  
\begin{align}
     \label{eq:LMnonzero}
    \Lag_{\rm eff}  &=  f_{\phi}^2 (\partial_{\mu} \theta_{\phi} \partial^{\mu} \theta_{\phi})  
    + f_{\eta}^2   (\partial_{\mu} \theta_{\eta} \partial^{\mu} \theta_{\eta})\\
    &\nonumber+ \Lambda^4  \cos\left (\theta_{\phi} +  \theta_{\eta} 
    -  \bar{\theta} \right )  +    \Lambda_{m}^4 \cos\left( \theta_{\eta} \right).
\end{align} 
 
First of all, we notice that all $CP$-violating phases are fully canceled in the vacuum, which is achieved for   
\begin{equation} 
    \theta_{\eta} =  0\, , ~~~\, \theta_{\phi} = \bar{\theta} \,.
\end{equation}  
Due to the last term in \eqref{eq:LMnonzero}, the previous mass eigenstates are slightly re-arranged. The correction to the eigenstates \eqref{eq:aeta} and \eqref{eq:aphi} are $\sim \mathcal{O}(m_\psi/\Lambda)$. So they are still given mostly by $\eta'$ and $\phi$ respectively. 

Additionally, the masses of the eigenstates $a_{\eta}$ and $a_{\phi}$ change. The mass of $a_\eta$ gets a correction 
\begin{equation} \label{eq:shiftMeta} 
    \delta m_{\eta}^2 \simeq \frac{\Lambda_m^4}{f_{\eta}^2} \sim m_{\psi}\Lambda \,. 
\end{equation}
More importantly, the previously zero-mass eigenstate $a_\phi$ now gains the mass 
\begin{equation} %\label{Mphi}
    m_{\phi} \simeq \frac{\Lambda_{m}^2}{f_{\phi}}\,.    
\end{equation} 
This is expected, since both $U(1)_V$ and $U(1)_A$ are now explicitly broken: $U(1)_A$ by the anomaly and  $U(1)_V$ by the mass term.

The true axion, i.e. the mode that makes $\theta_{\rm eff} = 0$, is now $\theta_\phi \sim a_\phi + \xi a_\eta$.
Thus, the true axion now resides predominantly in the $a_\phi$-boson with a small admixture from $a_\eta$.    
This boson is what traditionally is called the axion, however this terminology hides the essence of the story. 

Without $\phi$, the $\eta'$ would reduce the value of $\bar{\theta}_{\rm eff}$ down to a fraction $ \sim m_{\psi}/\Lambda$, as shown in Eq. \eqref{eq:etapotential}.
Thus, one can say that the $\eta'$-meson is already doing part of the job. 
However, this is not nearly sufficient for reducing the $ \bar{\theta}_{\rm eff}$ to the phenomenologically acceptable value.
Therefore, for solving the strong-$CP$ puzzle, having $\phi$ is crucial.

\section{Strings and walls} 
  
We shall now discuss the topological defects that arise in our model that takes into account the mixing between  $\eta'$ and the axion.
Since by shifting the fields we can eliminate the $CP$-violating phase,  without loss of generality we set $\bar{\theta} = 0$. In our analysis we repeat the same steps as before. Namely, we first find the topologically non-trivial solutions for $m_{\psi} =0$ and then analyze how these structures are affected by $m_{\psi} \neq 0$. 
    
\subsection{Massless quark case} 
    
We start with the case $m_{\psi} =0$. The effective Lagrangian of the phase degrees of freedom is given in Eq. \eqref{eq:LGoldMzero}. In the absence of the anomaly, both spontaneously broken symmetries $U(1)_A$ and $U(1)_V$ would be exact, with the topology of the vacuum manifold being  the two-torus $T_2 = S_1\times S_1$. Such a vacuum would support two independent integer winding numbers, $n_{\phi}$ and $n_{\eta}$, in the phases of the two condensates, $\langle \Phi \rangle$ and $\langle \bar{\psi} \psi \rangle$. Correspondingly, there would exist two types of cosmic strings with uncorrelated windings in $\theta_{\phi}$ and $\theta_{\eta}$ respectively. 

In polar coordinates, $r, \varphi, z$, around straight infinite strings along the $z$-axis the respective phases wind as
\begin{equation} %\label{windings1} 
    \theta_{\eta} = n_{\eta} \varphi\,, \, \  \theta_{\phi} = n_{\phi} \varphi\,. 
\end{equation}
The lowest energy configurations are given by winding numbers $|n_\phi| = |n_\eta| =1$.  
     
However, the QCD anomaly changes the picture and correlates the two windings. The main term responsible for the correlation is the instanton-induced potential. The structure can be understood by analyzing the effective low energy theory of the phase degrees of freedom.  

The equations of motion are: 
\begin{align}
    \square \theta_{\phi}   +  \frac{\Lambda^4}{2f^2_\phi} \,   \sin\left( \theta_{\phi} +  \theta_{\eta} \right) &= 0. \\ \label{eq:eomtheta}
    \square \theta_{\eta}  + \frac{\Lambda^4}{2f^2_\eta}  \, \sin\left( \theta_{\phi} +  \theta_{\eta} \right) &= 0. 
\end{align}
There exist  two classes of relevant solutions with correlated phases.  

\subsubsection{axion-$\eta'$ string of global $U(1)_V$}

First, we focus on the $U(1)_V$-symmetry.
As explained above, this is the $U(1)$ symmetry which is not explicitly but only spontaneously broken.
Thus, there exist global strings of $U(1)_V$.
Far away from the straight string the two phases wind oppositely: 
\begin{equation} 
    \theta_{\phi} =  - \theta_{\eta}  = n_V \varphi \, ,
\end{equation}
where $\varphi$ is the polar angle in the plane perpendicular to the string and $n_V$ is the winding number. The lowest energy string is the one with unit winding number, $n_V = \pm 1$. Due to the instanton-induced potential, there exists no string around which only one phase winds. 
 
This can be seen by evaluating the energy functional around a circle of radius $R$ in the perpendicular plane: 
\begin{equation} 
    \sigma =  \int_0^{2\pi R} dl \, ~
    f_{\phi}^2 (\partial_{l} \theta_{\phi})^2
    + f_{\eta}^2  (\partial_{l} \theta_{\eta})^2 
    - \Lambda^4  \cos\left (\theta_{\phi} +  \theta_{\eta}  \right ),
\end{equation} 
where $l = R\varphi$. The uniform winding of only one phase would necessarily create a non-uniformity (domain wall) due to the last term.  

In other words, a string solution without domain walls must not be explicitly broken by the cosine potential.
Thus, we need a field configuration with $\theta_\phi + \theta_\eta=0$, which requires that $\theta_\phi$ and $\theta_\eta$ wind oppositely.

\subsubsection{$\eta'$ string-wall system of $U(1)_A$}
Let's now turn to the anomalous $U(1)_A$ symmetry. This symmetry is spontaneously broken and, due to the anomaly, it is also explicitly broken. Thus, we have $U(1)_A$ strings attached to domain walls.

The domain wall appears because of the explicit breaking, i.e. the cosine potential. Energetically, the phase prefers to stay in the minimum as much as possible and concentrate the entire winding in a small region, which is the domain wall.

Naively, around a $U(1)_A$-string both phases would wind equally, $\theta_\phi=\theta_\eta = n_A \varphi$.
However, this is not favorable energetically. 
Instead, to leading order in the $f_\eta/f_\phi$ expansion, the lowest energy string winds only in the $\theta_\eta$-phase. This winding corresponds to a 
transformation that combines $U(1)_A$ and $U(1)_V$.

The wall of $\eta'$, with the wall placed at $l=0$ in the compact coordinate, is described by\footnote{This is the solution to the eom in Eq. \eqref{eq:eomtheta} above, which is the Sine-Gordon Equation. The solution is for an infinite radius and represents a good approximation for $R \to \infty$.}
\begin{equation} \label{eq:Weta1}
    \theta_{\eta} = 4 \arctan( {\rm e}^{m_{\eta} l})\,, \ \  
    \theta_{\phi}  = 0 \,,
\end{equation} 
where the mass $m_\eta$ was defined in Eq. \eqref{eq:masseta}.

The tension (energy per unit area) of the $\eta'$-wall is given by 
\begin{equation} %\label{Teta1}
    \sigma_{\eta} \sim \Lambda^2 f_\eta.
\end{equation}  
If instead of winding in $\theta_{\eta}$ we wind in  $\theta_{\phi}$, the corresponding domain wall would have a much higher tension
\begin{equation} %\label{Teta}
    \sigma_{\phi} \sim  \Lambda^2 f_{\phi}  \gg \sigma_{\eta}  \,.
\end{equation} 
Therefore the energy cost of these domain walls is much larger. 
  
To conclude, in the lowest energy 
topological defects
of both types (strings as well as walls bounded by strings) the phase of the light 
quark condensate, $\theta_{\eta}$, winds.  Correspondingly, in the core of the strings the quark condensate must vanish.

\subsection{Massive light quark}    
   
We now deform the theory by putting $m_{\psi} \neq 0$. The effective Lagrangian is given by Eq. \eqref{eq:LMnonzero}. The last term comes from the quark mass that breaks the $U(1)_V$-symmetry explicitly. 
In the following, we consider the effect of this explicit breaking on the structure of the topological defects.

\subsubsection{$\eta'$ string wall system} 
  
For $m_{\psi} \ll \Lambda$, the effect of a light quark mass term  on the $U(1)_A$ string-wall system is insignificant, and it is still dominated by the winding of the $\theta_{\eta}$-phase. In the first approximation, the solution is still given by Eq. \eqref{eq:Weta1} with the mass $m_{\eta}$ slightly shifted by the explicit breaking term as given by Eq. \eqref{eq:shiftMeta}.

\subsubsection{$\eta'$-axion string wall system} 

The effect of $m_{\psi} \neq 0$ on the would-be $U(1)_V$-string is more dramatic, since a domain wall gets attached to it.  However, the phase $\theta_{\eta}$ continues to fully wind around the string. 

This can be understood in simple terms by analyzing the energy balance among various terms in the energy functional around a large circle of radius $R \gg m_{\phi}^{-1}$. After the inclusion of the mass term it becomes
\begin{align}  \label{eq:EwithM}
    \sigma &=  \int_0^{2\pi R}  dl \, ~
    f_{\phi}^2\, (\partial_{l} \theta_{\phi})^2
    + f_{\eta}^2 \,  (\partial_{l} \theta_{\eta})^2+\\
    &\nonumber  \Lambda^4\left(1- \cos\left (\theta_{\phi} +  \theta_{\eta}  \right ) \right )
    \, + \,  \Lambda_m^4\left (1-  \cos\left (\theta_{\eta} \right ) \right ),
\end{align} 
where for convenience we normalized the energy of the vacuum to zero. 
Since the topological structure is determined by the asymptotic values of the fields far away from the string core, for large $R$, the variational problem is effectively one-dimensional. 
  
Let us examine what is traditionally called an axionic string. This string is usually defined as the one with non-zero winding number  $n_{\phi}$ of the Peccei-Quinn field phase  $\theta_{\phi}$. The uniform winding of only  this  phase, $\theta_{\phi} = l/R$, would cost energy $\sigma_{R} = 2\pi (f_{\phi}^2/R +  R\Lambda^4)$, which diverges linearly with $R$. 
   
The system can lower this divergent energy by two tricks:  
  
{\it 1)} The opposite winding in $\theta_{\eta} = -\theta_{\phi}$, which gets rid of the energy of the $\Lambda^4$-term. Since $\Lambda \gg \Lambda_m$, the first cosine in Eq. \eqref{eq:EwithM} is much more costly in energy; 

{\it 2)}  Limiting the non-trivial winding of the phases to fixed arches, while keeping the rest of the circle in the vacuum. This turns the axionic string into a string-wall system.
\begin{figure}
    \centering
    \includegraphics[width=0.5\linewidth]{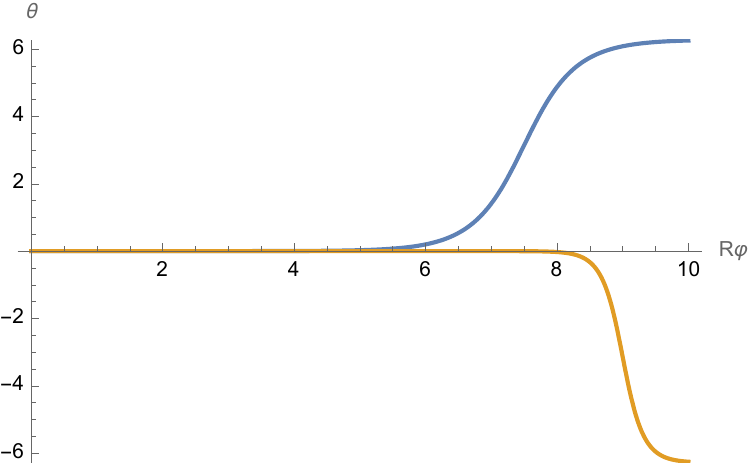}
    \includegraphics[width=0.5\linewidth]{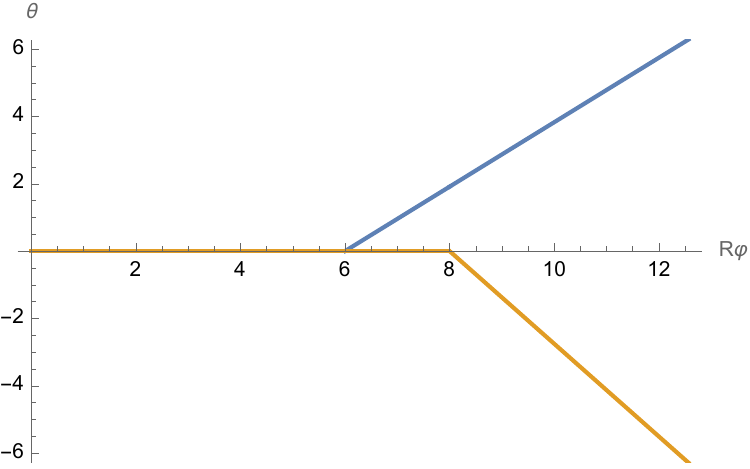}
    \caption{Illustration of the wall profiles for the linear ansatz (\ref{eq:ansatzeta}), (\ref{eq:ansatzphi}) and for their smoothed out versions. This behavior is confirmed by numerical analysis \cite{numerics}. }
        \label{fig:phase_winding_linear}
\end{figure}
  
The winding length can be estimated by balancing the gradient energy and the potential energy. 
To leading order, we approximate the windings with a linear ansatz:
\begin{equation} \label{eq:ansatzeta}
    \theta_{\eta}  =  \frac{l}{L_{\eta}} \quad {\rm for} \quad   0 \leqslant l \leqslant 2\pi L_{\eta}       
 \end{equation} 
and
\begin{equation} \label{eq:ansatzphi}
    \theta_{\phi} =  -  \frac{l}{L_{\phi}} \quad {\rm for} \quad   0 \leqslant l \leqslant 2\pi L_{\phi } ,
\end{equation}
while outside of the respective arches the phases stay constant and in the vacuum.
The $L_\eta$ and $L_\phi$ are the lengths where the phases wind.
They are also the domain wall thickness.
We expect $L_\phi \geq L_\eta$ because $f_\phi > f_\eta$.
The functions are shown in \autoref{fig:phase_winding_linear}.

 By explicit minimization of the functional 
 (\ref{eq:EwithM}) (see the 
\hyperref[sec:appendix]{Appendix}), we estimate that the energetically most favorable regime is $L_{\phi} = L_{\eta} \equiv  L$ with the optimal winding length  
\begin{equation} \label{eq:L1}
    L  = \frac{f}{\Lambda_m^2},
\end{equation}  
where we used $f \equiv \sqrt{f_\phi^2 + f_{\eta}^2}$.
The length of the winding arch \eqref{eq:L1} sets the thickness of the domain wall: 
\begin{equation}    
   L_* =  \frac{f}{\Lambda^2_m}  \sim m_{\phi}^{-1},
\end{equation}
which matches the standard thickness of the axionic domain wall. The novelty is the simultaneous winding of~$\theta_{\eta}$.  
   
We can refine the above energetic estimate by imposing the constraint $\theta_{\phi} = -\theta_{\eta}$ on the energy functional \eqref{eq:EwithM} and solving the equation of motion. For $R \gg m_{\phi}^{-1}$ this gives,      
\begin{equation} \label{Lgold}
    \theta_{\phi}  \simeq -  \theta_{\eta} \simeq 4 \arctan( {\rm e}^{m_{\phi} l})\,. 
\end{equation} 
Of course, in reality, the above wall configuration  has to be understood as the leading order estimate in an $m_{\psi}/\Lambda$ expansion.   

We conclude that, in a theory in which the PQ field communicates to QCD via a heavy quark, what is traditionally called the axion string-wall system
involves winding of the phase of the light quark condensate, involving the $\eta'$-meson. This has some important implications.

\subsection{Implications for Anomaly inflow} 

It is well known that fermions, which receive their mass from Yukawa couplings to the string field (such as the Peccei-Quinn field  $\Phi$) with topologically non-trivial phase, deposit zero modes on the string \cite{Jackiw:1981ee}.  
 
From the point of view of the $1+1$-dimensional theory on the string world-sheet these are massless fermions. Moreover, each Dirac flavor coupled to $\Phi$ with winding number $n$ deposits $n$ zero modes of definite chirality, determined by the sign of the winding number.   
    
Witten showed \cite{Witten:1984eb} that when the fermions are electrically charged, their zero modes can carry a superconducting current along the string. That is, the string behaves as sort of a superconducting wire. This has important astrophysical implications
\cite{Ostriker:1986xc}. 

When the global symmetry forming a string is anomalous, the charges of the localized fermions are not balanced. In such a case, despite the fact that the parent $3+1$-dimensional theory has no gauge anomaly, the $1+1$ dimensional theory of zero modes appears to be anomalous. This question was  addressed by  Callan and Harvey, who showed that the anomaly is canceled by the inflow of charge from the bulk onto the string \cite{Callan:1984sa}. 

It was discussed by Lazarides and Shafi that such a phenomenon must be exhibited by axionic strings, since the global symmetry forming it is (assumed to be) the anomalous Peccei-Quinn symmetry \cite{Lazarides:1984zq}. They suggested that the anomaly inflow must take place along the domain walls attached to the string (for more discussions, see \cite{Naculich:1987ci,Harvey:2000yg}).

In general, the presumed anomaly on the axionic string can have important astrophysical implications due to the correlation between the charge and chirality of the current-carriers  \cite{Agrawal:2019lkr,Agrawal:2020euj,Fukuda:2020kym}.

Our analysis sheds a novel light on the question of anomaly inflow on the axionic string. Namely, in the considered prototype model with one heavy and one light quark,
around the string of the lowest energy, the phase of the quark condensate, $\theta_{\eta}$,
winds oppositely to the phase of the Peccei-Quinn field $\Phi$. Due to this, the set of deposited zero modes is anomaly free. 
 
To see this, let us consider the UV-Lagrangian from Eq.~\eqref{eq:LPQ}. The phase of the PQ field $\theta_\phi$
winds around the axionic string. Since the heavy quark $\Psi$ has a Yukawa coupling with the PQ field, it deposits a zero mode. This zero mode has a definite chirality and it carries both electric charge as well as color, thereby creating an anomalous $1+1$-dimensional current. 
   
However, we have seen that, in the lowest energy state, the phase of the light quark condensate, $\theta_{\eta}$, winds in the opposite way. This has two effects. First, the quark condensate vanishes in the core of the string. Secondly, the light quark $\psi$ deposits a zero mode in the core of the string.

The appearance of the light fermion zero mode on the string can be seen in several ways. The shortcut is the index theorem. 
   
\begin{figure}
    \centering
        \begin{tikzpicture}
            \def\r{1}
            \begin{feynman}
                \vertex (a) at (0,0);
                \vertex (p) at (-\r,0){$\phi$};
                \vertex[blob] (t) at (2*\r,0){};
                \vertex (x1) at (3*\r,\r) {$\psi$};
                \vertex (x2) at (3*\r,-\r){$\psi$};

            \diagram*{
                (p) --[scalar] (a) --[fermion,half left, out=80, in=100, edge label = \(\Psi\)] (t)[crossed dot] --[anti fermion, half left, in=100, out=80, edge label = \(\Psi\)] (a), (x1) --[fermion] (t) [blob] --[anti fermion] (x2),};
            \end{feynman}
        \end{tikzpicture}
    \caption{A diagram describing how the effective Yukawa coupling \eqref{tHY} is generated by the 't Hooft vertex}
    \label{fig:yukawa}
\end{figure}
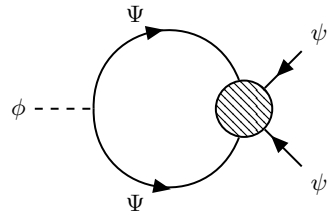

It can also be seen directly, by noticing that via the 't Hooft  determinant the instantons generate a direct Yukawa coupling between the Peccei-Quinn field $\Phi$ and the light fermions, 
\begin{equation} \label{tHY} 
    \Phi \bar{\psi}_L\psi_R \, + \, {\rm h.c}.
\end{equation} 
The diagram generating this effective interaction is shown in \autoref{fig:yukawa}.
This coupling breaks $U(1)_A$ but it is invariant under the anomaly-free $U(1)_V$ symmetry. 
Hence, the Peccei-Quinn string with winding number $n= +1$ is seen by the light fermions as the anti-string with winding number $n=-1$. Correspondingly, the localized zero mode from $\psi$ has the opposite chirality from the one deposited by the heavy fermion $\Psi$.

Since the zero modes of the $\psi$ and $\Psi$ have opposite chirality, their gauge anomalies cancel out. 
As a result, the fermion content of $1+1$-dimensional theory on the string is anomaly free.   

This is also apparent from the fact that around the minimal string the VEVs wind by the $U(1)_V$-transformation, which is an anomaly-free symmetry. 

% \footnote{For axionic models involving more flavors (see below), the lowest energy string does not have to wind around the anomaly free $U(1)_V$. Thus, in general, 
% depending on the field content and the winding numbers,
% the axionic string can support an anomalous current.}. 

The explicit mass term $m_{\psi}$ of the light quark does not affect the zero mode structure as long as $m_{\psi} \ll \Lambda$.
This follows from the general consideration of the effect of an explicit bulk mass of the fermion on the zero mode structure on the string \cite{Bagherian:2023jxy}. 

In general, if the mass gap generated by the explicit breaking of the chiral symmetry is smaller than the gap generated from the spontaneous breaking, the zero mode fermion still exists. The effect of explicit breaking boils down to a small deformation of the wave-function profile of the zero mode.

Of course, for  $m_{\psi} \neq 0$, simultaneously, domain walls get attached to the string, as already discussed. The thickness of the wall is much larger than the size of the string core.

As explained in the previous section, there exists a minimal $\eta'$ string which winds only in $\eta'$.
This winding involves the anomalous $U(1)_A$ symmetry such that the $\eta'$ string has an anomaly.

The general message of this section is that the winding of the phase of the quark condensate around the string changes its zero mode structure and, correspondingly, the anomaly content. In particular, even for a string formed by the PQ field, the energetics can force the phases to wind by an anomaly-free transformation, implying that the zero mode content of the axion string is anomaly free. 
In general, the anomaly content of the string depends both on the quark flavor structure of the model and on the specifics of their winding numbers.

\subsection{More heavy quark flavors} 
  
Let us now discuss how the existence of multiple heavy quark flavors  $\Psi_i$ changes the story. The index $i$ runs over the number of flavors $N_f$. At the same time, we keep a single light quark flavor.

The anomaly-free symmetry $U(1)_V$ acts on the fields as  
\begin{equation} %\label{U1VN}
    \Phi \rightarrow {\rm e}^{i\alpha} \Phi\,,\quad \Psi_i \rightarrow {\rm e}^{i\frac{1}{2}\alpha \gamma_5} \Psi_i\,,\quad  \psi \rightarrow {\rm e}^{-i\frac{N_f}{2}\alpha \gamma_5} \psi\,.
\end{equation} 
  
At the same time, under the anomalous $U(1)_A$-symmetry the fields transform as  
\begin{equation} %\label{U1AN}
    \Phi \rightarrow {\rm e}^{i\alpha} \Phi\,,\quad  
    \Psi_i \rightarrow {\rm e}^{i\frac{1}{2}\alpha \gamma_5} \Psi_i\,,\quad  \psi \rightarrow {\rm e}^{i\frac{1}{2}\alpha \gamma_5} \psi\,.
\end{equation} 
The $U(1)_A$ possesses an anomaly-free $Z_{2N_f}$-subgroup, for $\alpha = \frac{2\pi k}{N_f}$ with $k = 1,2,...,2N_f$.
 
This is also visible from the effective Lagrangian of the phases, which takes the form
\begin{align} \label{LMnonzero2}
    \Ll &=  f_{\phi}^2\, (\partial_{\mu} \theta_{\phi} \partial^{\mu} \theta_{\phi})  
    + f_{\eta}^2 \,  (\partial_{\mu} \theta_{\eta} \partial^{\mu} \theta_{\eta})+\\
    &\nonumber+  \, \Lambda^4\, \left(\cos\left (N_{f} \theta_{\phi} +  \theta_{\eta} 
    -  \bar{\theta} \right ) -1)\right )+\\
    &\nonumber+    \Lambda_{m}^4 \left( \cos\left( \theta_{\eta} \right) - 1)\right ) \,.
\end{align} 
For $\Lambda_{m} =0$, the Lagrangian is invariant under $U(1)_V$ as well as under $Z_{N_f}$ (the phases of the condensates do not transform under $Z_2$ that flips the signs of fermions). Again, without loss of generality we can set $\bar{\theta} = 0$. 
  
Let us now consider topological defects. Setting first $\Lambda_{m} =0$, we have two types of defects. The first type are $U(1)_V$-strings. Since the winding numbers must be integer, the string with minimal winding numbers has
\begin{equation} 
    n_{\eta} = \pm N_f \quad {\rm and}\quad n_{\phi} = \mp 1\,. 
\end{equation} 
Correspondingly, the phases are
\begin{equation}  \label{windingsV2} 
    \theta_{\eta} = \pm N_f \varphi\,, \quad \theta_{\phi} = \mp \varphi\,. 
\end{equation} 
After switching on the $\Lambda_m^4$-term, this string becomes a junction of $N_f$ domain walls. Across each wall, the changes of the phases are given by $\Delta  \theta_{\eta} =2\pi$ and $\Delta  \theta_{\phi} = \frac{2\pi}{N_f}$ respectively, whereas in between the walls the phases stay constant. 

The second type of defect is the $\eta'$-string wall system which behaves just as in the case of one heavy flavor. As before, it is energetically more favorable to only have winding in the $\eta'$ direction instead of both.

%    Again, in the leading order in $m_{\psi}/\Lambda$ expansion,  the wall solution is,     
%              \begin{equation} \label{Lgold}
%  \theta_{\phi}  \simeq -  \frac{1}{N_f} \theta_{\eta} \simeq 
%   \frac{4}{N_f} \arctan( {\rm e}^{x/L_*})\,. 
% \end{equation} 
% where, the characteristic thickness of the wall is given by 
% $L_* \simeq \Lambda_m^2/(f\sqrt{N_f}) \sim 1/m_{\phi}$.  
%     

\section{axion models}

The above-discussed simplified model with one heavy and one light quark captures the essence of the story for fully realistic axion models. 
  
For the heavy quark hidden axion model (also known as KSVZ model) \cite{Kim:1979if,*Shifman:1979if}, in which the PQ field couples to QCD via an exotic heavy quark, the connection is straightforward.
In our model, the $\Psi$ is the exotic heavy quark.
The role of the light quark $\psi$ is played by the up-quark, $u$, which is the lightest quark in the Standard Model. 
  
The same is true for scenarios of DFSZ type \cite{Dine:1981rt,*Zhitnitsky:1980tq}. In these models no exotic quarks are introduced. Instead, the PQ field communicates with the QCD sector through the ordinary quarks via Higgs fields. For illustrative purposes, it suffices to consider a single generation realization of this scenario, in which we consider two flavors of light quarks, $u$ and $d$.  
   
To make the chiral $U(1)_{PQ}$ symmetry compatible with non-zero quark masses, one needs to introduce two designated Higgs doublets that give masses to the $u$ and $d$ quarks respectively.  We shall focus on the  components with non-zero VEVs  which are electrically neutral and we denote them by $H_u(x) = \rho_u(x) e^{i\chi_u(x)}$ and $H_d(x) = \rho_d (x) e^{i\chi_d(x)}$\footnote{The electrically-charged  components of the doublets are not relevant for the quark masses and play no role in the current discussion.
}. The quark masses are generated from the following Yukawa couplings, 
\begin{equation} \label{YHdHu}
    g_u H_u \bar{u}_Lu_R  \, + \, g_d H_d \bar{d}_Ld_R \, + 
    {\rm h.c.} ,
\end{equation} 
where $g_u$ and $g_d$ are the Yukawa coupling constants. $H_u$ and $H_d$ carry opposite gauge hypercharges. The above Yukawa couplings are invariant under the chiral PQ symmetry, 
\begin{align} \label{UPQ}
   &H_u \rightarrow {\rm e}^{i\alpha} H_u\,,\quad 
   H_d \rightarrow {\rm e}^{i\alpha} H_d\,,\\   
   \nonumber &u \rightarrow {\rm e}^{-i\frac{1}{2}\alpha \gamma_5} u\,,
  \quad d \rightarrow {\rm e}^{-i\frac{1}{2}\alpha \gamma_5} d\,.
\end{align} 
The non-zero VEVs $\langle \rho_u \rangle = v_u,\ \langle \rho_d \rangle = v_d$  spontaneously break the $U(1)_{PQ}$-symmetry and generate the quark masses, $m_u = g_u v_u,\  m_d=g_dv_d$. One combination of the phase degrees of freedom becomes a longitudinal component of the electroweak $Z$-boson, whereas the other would be an axion. However, it is well-known that since the VEVs  $v_u, v_d$ are bound from above by the weak interaction scale, such an axion is phenomenologically inconsistent, as its interaction with the Standard Model would be too strong. In order to ``hide" it, it is necessary to mix  $\chi_u, \chi_d$ with the phase of a PQ field, $\theta_{\phi}$, with much higher scale. This is achieved through the following phase-dependent coupling in the Lagrangian, 
\begin{equation} \label{mixing}  
    \mu \, \Phi\, H_u\, H_d  \, + \, {\rm h.c.} \,.
\end{equation} 
Under the PQ symmetry, $\Phi$ transforms as 
\begin{equation} \label{PhiPQ}
   \Phi  \rightarrow {\rm e}^{-i2\alpha} \Phi \,.
\end{equation} 
    
Below the QCD scale quarks condense. Since the quark masses are different, it is useful to keep track of their phase degrees of freedom separately: $\langle \bar{u}u \rangle = \Lambda_u^3 e^{i\theta_u(x)}, \ \langle \bar{d}d \rangle = \Lambda_d^3 e^{i\theta_d(x)}$, with scales $\Lambda_u \simeq \Lambda_d \sim \Lambda$. 
   
After taking into account instanton effects, the effective Lagrangian for phase degrees of freedom becomes
\begin{align} \label{Kphi}
 \Lag_{\rm eff} & = f_{\phi}^2\, (\partial_{\mu} \theta_{\phi})^2   + \\
 &+ v_{u}^2 \, (\partial_{\mu} \chi_{u} - g Z_{\mu})^2 +  v_{d}^2 \,  (\partial_{\mu} \chi_{d}  + g Z_{\mu})^2 \, + \label{Kchi} \\
 &+ f_{u}^2 \,  (\partial_{\mu} \theta_{u}  + g Z_{\mu})^2 +  f_{d}^2 \,  (\partial_{\mu} \theta_{d} - g Z_{\mu})^2 + \label{Ktheta} \\ 
 &+ \mu f_{\phi} v_u v_d\,  \cos\left (\theta_{\phi} +  \chi_u + \chi_d \right ) +  \label{PhiHH}\\
 &+ m_u\Lambda_u^3  \cos\left (\chi_{u} +  \theta_{u} \right ) + m_d\Lambda_d^3  \cos\left (\chi_{d} +  \theta_{d} \right ) +  \label{Ycond} \\
 &+ m_d\Lambda_u^3\,  \cos\left ( \theta_{u}  - \chi_{d}
  -  \bar{\theta} \right ) + \label{Yinst1}\\
  &+ m_u\Lambda_d^3\,  \cos\left ( \theta_{d}  - \chi_{u}
  -  \bar{\theta} \right ) + \label{Yinst2}\\
  & + m_um_d\Lambda^2\,  \cos\left (\chi_{u} + \chi_d 
  +  \bar{\theta} \right ) +  \label{Yinst3} \\
  & + \Lambda^4\,  \cos\left (\theta_{u} +  \theta_{d} 
  -  \bar{\theta} \right )  \label{Inst} \,.  
 \end{align}

Let us make a few clarifying remarks.  The first line represents the kinetic term of the phase of the PQ field. This field carries no gauge quantum number and does not couple directly to any of the gauge fields or fermions of the Standard Model.  

The lines (\ref{Kchi}) and (\ref{Ktheta}) describe the kinetic terms of the phases of the Higgses and the quark condensates respectively. All these phases shift under the combination of the gauge hypercharge and weak isospin symmetry of the Standard Model, which acts on the $Z$-boson as
\begin{equation} \label{gaugeZ}
    Z_{\mu} \rightarrow Z_{\mu}  + \partial_{\mu} \beta \,,
\end{equation} 
where $\beta(x)$ is the gauge parameter. Under this symmetry the phases shift as
\begin{equation} \label{gaugePhases}
    \chi_{u} \rightarrow   \chi_{u}  +  \beta\,, ~
    \chi_{d}  \rightarrow   \chi_{d}  -  \beta\,, ~ 
    \theta_{u}  \rightarrow   \theta_{u}  -  \beta\,, ~  
    \theta_{d}  \rightarrow   \theta_{d}  +  \beta \,.
\end{equation} 
That is, the gauge symmetry \eqref{gaugeZ} is Higgsed not only by the VEVs of the Higgs doublets, but also by the quark condensates. Correspondingly, the following combination of the phases,
\begin{equation} \label{phaseZ}
  v_{u}^2  \chi_{u}-  v_{d}^2  \chi_{d}  - f_{u}^2 \,  \theta_{u}  +  f_{d}^2  \theta_{d} \,,
\end{equation} 
becomes the longitudinal component of the $Z$-boson. 
  
Line (\ref{PhiHH}) corresponds to the phase-dependent potential term emerging from the cross coupling (\ref{mixing}) between PQ and Higgs fields. Line (\ref{Ycond}) describes the phase-dependent potential emerging from the Yukawa couplings between the Higgses and the respective quark condensates in equation \eqref{YHdHu}.  

The couplings \eqref{Yinst1}, \eqref{Yinst2} and \eqref{Yinst3} describe the terms generated by combining various Yukawa couplings with the 't Hooft determinant. Finally, the term (\ref{Inst}) is the quark-mass-independent instanton contribution.   
 
As it is clear from Eq. \eqref{gaugePhases}, all the arguments of the phase-dependent potentials are gauge invariant, as they should be. It is clear that in the minimum of the potential all the $CP$-odd phases vanish and the strong-$CP$ problem is solved. Therefore, without loss of generality, we can set $\bar{\theta}=0$. 
   
Let us now discuss the structure of cosmic strings and domain walls arising in this scenario. First, notice that while the axionic strings are expected to be global, they nevertheless carry a gauge magnetic flux  \cite{Dvali:1993sg}. This may come as a surprise, since the naive expectation is that around axion-like strings the phases wind by the global $U(1)_{PQ}$-transformation.  
  
However, the requirement that each phase must be single-valued modulo $2\pi$, implies that the topologically non-trivial background sources the gauge field. That is, around any string the winding number of each phase must be an integer. As shown in \cite{Dvali:1993sg} this implies that the axionic string carries a magnetic flux of the $Z$-boson. This flux is given by 
\begin{equation} \label{Zmagnet}
    \mathrm{flux} = \oint dx_{\mu} Z^{\mu} = 
    \frac{1}{g_z} \frac{v_{u}^2  n_{\chi_{u}}
    -  v_{d}^2  n_{\chi_{d}}  - 
    f_{u}^2 \,  n_{\theta_{u}}  +  f_{d}^2  n_{\theta_{d}}}{
    v_{u}^2  +  v_{d}^2  +  
    f_{u}^2 \,  +  f_{d}^2 } \,,
\end{equation}
where the integral is taken around an asymptotic closed 
path around the string. This flux is non-integer in units of the inverse gauge coupling of the $Z$-boson, 
$1/g_z$. Thus, unlike the case of the Nielsen-Olesen string, the asymptotic pure-gauge form of the gauge field cannot entirely compensate the gradient energy of the winding phases. Correspondingly, the energy per unit length of the string is logarithmically divergent in the transverse direction. Thus, the axionic string shares properties of both local and global strings. Because of this, in \cite{Dvali:1993sg} such strings were referred to as ``semi-global". 
 
Notice that due to the hierarchy between the weak and QCD scales, $v_{u}, v_{d} \gg  f_{u}, f_{d} $, the contribution to the $Z$-flux from the phases of the quark condensate relative to the phases of the Higgs bosons are negligible ($\sim  10^{-4}$) and can be ignored. 
Therefore, the flux can be well-approximated by
\begin{equation} \label{ZfluxH1}
    \mathrm{flux}  \simeq 
    \frac{1}{g_z} \left (n_{\chi_{u}} - \frac{v_{d}^2  (n_{\chi_{u}}
    + n_{\chi_{d}})  }{
    v_{u}^2  +  v_{d}^2  }  \right )\,. 
\end{equation}
Notice that an accidental cancellation of the flux, due to an interplay between the VEVs and the winding numbers, is highly unlikely. Specifically, in the case of minimal winding, e.g. $n_{\chi_{u}} = \pm 1, \ n_{\chi_{d}} = 0$, the flux would be
\begin{align}\label{minflux}
    \mathrm{flux} \simeq \pm\frac{1}{g_z}\frac{v_u^2}{v_d^2+v_u^2},
\end{align}
and the cancellation of the flux is impossible even at the expense of fine tuning.

\subsection{Massless up quark case}
Let us now discuss various winding configurations that give strings and walls. As in the previous sections, we start by setting the mass of the lightest quark, the up quark, to zero. For $m_u=0$, the effective potential is 
\begin{align} \label{muzero}
V &=  
 - \mu f_{\phi} v_u v_d\,  \cos\left (\theta_{\phi} +  \chi_u 
 + \chi_d \right ) -\\
  \nonumber &- m_d\Lambda_d^3\,  \cos\left (\chi_{d} +  \theta_{d} \right ) -   
  \nonumber 
 m_d\Lambda_u^3\,  \cos\left ( \theta_{u}  - \chi_{d} \right ) - \\ 
 \nonumber &-\Lambda^4\,  \cos\left (\theta_{u} +  \theta_{d} \right ) \,.  
\end{align}
This potential is invariant under the following 
anomaly free global $U(1)_V$-symmetry,
\begin{align} \label{exactU1}
    &\theta_{\phi}  \rightarrow   \theta_{\phi}  -  2\alpha \,,\ 
    \chi_{u} \rightarrow   \chi_{u}  +  \alpha\,, \  
    \chi_{d}  \rightarrow   \chi_{d}  +  \alpha \,, \\ \nonumber  
    &\theta_{u}  \rightarrow   \theta_{u}  +  \alpha\,, \   
    \theta_{d}  \rightarrow   \theta_{d}  -  \alpha \,.
\end{align} 
Notice that under this symmetry $\theta_{u}$ and $\theta_{d}$ shift in the opposite way. Since the symmetry is non-linearly realized, there exists a corresponding massless Nambu-Goldstone boson, which resides predominantly in  the combination $2\theta_{\phi}-\chi_u-\chi_d$\footnote{This is the combination that is not getting a mass from the term \eqref{mixing}. In terms of canonically normalized fields, this combination is predominantly made out of $\phi$, due to the hierarchy $f_\phi \gg v_u,v_d$.}, with a small admixture from the neutral pion, $\pi^0$.
% After turning on the up quark mass, this will become the axion. 
At the same time, the axion that cancels $\bar{\theta}$ is the $\eta'$-meson residing in $\theta_{u} +  \theta_{d}$.  
     
Since the symmetry  (\ref{exactU1}) is exact, it can give rise to anomaly-free PQ strings, similar 
to the ones we discussed previously. 

Consider a string with unit winding numbers, $n_{\phi} = \pm 1$. In order to keep the potential \eqref{muzero} vanishing while retaining the single-valued phases, only one of the Higgs phases can wind around the string. As explained, in order to reconcile the topological non-triviality with single valued phases, winding cannot be purely by a global $U(1)_V$ transformation but must be combined with a $U(1)_Z$-gauge transformation (\ref{gaugePhases}). 
 According to \cite{Dvali:1993sg}, this leads to a nonvanishing magnetic $Z$-flux in the core of the string, given by  (\ref{Zmagnet}) (or approximately 
 by (\ref{ZfluxH1})).
 
With only one of the Higgs phases winding, there exist two winding options: $n_{\chi_{u}} = \mp 1,~ n_{\chi_{d}} =0$ or vice versa $n_{\chi_{u}} = 0,~ n_{\chi_{d}} = \mp 1$. In both cases there is a $Z$-boson magnetic flux trapped in the string with the value (\ref{ZfluxH1}).  
 
In the first case, the quark condensates do not wind $n_{\theta_{u}} = n_{\theta_{d}} = 0$. In the second case, the winding numbers are, $n_{\phi} = - n_{\chi_{d}} =  - n_{\theta_{u}} =  n_{\theta_{d}}$, in order to keep the potential vanishing. This winding
pattern of the quark phases around the string corresponds to a neutral pion, $\pi^0$.

Analogously, we can consider strings with higher winding numbers. 
All configurations in which $\chi_{d}$ winds are accompanied by windings of the 
quark condensates with the winding numbers satisfying the relation
$n_{\chi_{d}} =   n_{\theta_{u}} = - n_{\theta_{d}}$.

It is instructive to compare this to our discussion of the $U(1)_V$ PQ string in the
KSVZ-type model  (\ref{LUV}) of the previous section. 
There, in order to keep the potential vanishing, it was necessary for $\theta_{\eta}$ to wind around the string oppositely to $\theta_\phi$.
 In the present case of a DFSZ-type model, the situation is different, as the PQ field does not directly couple to the quarks. The winding of $\theta_\phi$ is compensated by the Higgs phases. 
 Therefore, as long as one of the quarks is massless, 
 depending on which of the Higgs phases winds, 
the winding of the quark phases may or may not be necessary.  
  However, when necessary,  both phases 
  wind simultaneously in the opposite directions,
  $\theta_u = -\theta_d$, thereby, realizing the 
  $\pi^0$ winding pattern in the flavor space 
 \footnote{In fact, since the quark phases contribute to the overall energy via the kinetic terms and the gauge flux, the case of $\chi_u$ winding is slightly favored compared to $\chi_d$ winding. However, the contribution from the quark phases is $\sim f_u^2-f_d^2$, which is negligible. Similarly, the contribution from the flux is $\sim f_u^2+f_d^2$.}. 

\subsection{Massive up quark case}
Now we wish to investigate what happens to the $U(1)_V$ strings, once we switch on $m_u\neq 0$. This mass induces an explicit breaking of the $U(1)_V$-symmetry  via the following additional terms in the effective potential
\begin{align}\label{VmDFSZ}
    V_{m_u} &\equiv -m_u\Lambda_u^3  \cos\left (\chi_{u} +  \theta_{u} \right ) - m_u\Lambda_d^3\,  \cos\left ( \theta_{d}  - \chi_{u}
    \right )- \\ \nonumber
    &- m_um_d\Lambda^2\,  \cos\left (\chi_{u} + \chi_d 
    \right ).
\end{align}
The explicit breaking gives rise to domain walls, 
which get attached to the strings. 

Let us take another look at the situation with $n_\phi =1$. In the massless 
up-quark case we found that, in order to cancel the winding of the PQ phase,
at least one of the Higgs phases must wind.
 For $n_{\chi_d} = 0$ 
(at least in the lowest energy configuration) there was no winding necessary in the quark phases, whereas the case $n_{\chi_d} \neq 0$ would induce winding in $\pi^0$. 

In the massive case, due to the additional terms
(\ref{VmDFSZ}) induced by $m_u$, the winding in $\chi_u$ can stimulate additional windings (or non-trivial angular dependences) in the quark phases. However, in a sector with given $n_{\chi_u}$ the ground state configuration  is determined by a more detailed energy balance.

% \footnote{There will also be winding induced in the phase $\chi_d$ of the down Higgs. However, this winding should be suppressed by a factor $\sim m_u/\Lambda$ {\bf (THIS LAST SENTENCE IS UNCLEAR, SINCE WINDING IS AN INTEGER. I think  if the cross term between PQ and Higgs fields is the dominant one (as it would be natural)
% the system will prefer not to wind the second Higgs. But perhaps we leave this out??}}. 

We observe that, just as in the case of KSVZ-type models 
with heavy exotic quarks, also in DFSZ-like theories
the winding in the PQ axion is in general accompanied by winding in the phases of quark condensate. 
%For a more detailed numerical analysis we refer to \cite{numerics}

\section{Internal structure of QCD walls} 
   
As we have seen, the topological structure of the  QCD vacuum supports the existence of  $2\pi$ domain walls. Across the wall, the phase of the quark condensate winds by $2\pi$.
The winding  can assume different patterns of flavor-dependence. Focusing on the quark condensates of the first generation,  $\langle \bar{u}u \rangle $ and $\langle \bar{d}d \rangle $, there exist windings in two basic flavor directions.    

In the case $n_{\theta_u} = - n_{\theta_d} \neq 0$, we can say that the winding takes place in the direction of $\pi^0$. In the case of  $n_{\theta_u} + n_{\theta_d} \neq 0$, we can say that the winding involves the $\eta'$-meson. 
       
In both cases the domain wall can come in the form of a closed bubble of a certain radius, or as a membrane bounded by a string. A planar (infinite) wall can also exist. However, since the vacua on the both sides of the wall are identical, a hole can be punched through it. The edge of this hole represents a loop of a cosmic string. Thus, for each category 
($\pi^0$ or $\eta'$) the existence a $2\pi$ domain wall implies the existence of a corresponding string that bounds it.  
   
Let us now discuss the internal structure of these walls. The important question that we would like to address is whether the order parameter in form of the quark condensate remains non-zero throughout the wall. This is a necessary condition for the phase to be well-defined.  When we pass through the wall, the quark condensate can  rotate in the flavor space non-trivially. As long as its absolute value stays non-zero, the winding of the phase is well defined and the existence of the wall is guaranteed by topology. 
 
Notice that the answer to this question is important for $\eta'$ domain walls formed by the quark condensate in pure QCD, without an external axion. For axionic domain walls of the PQ symmetry, the question is less pressing, since the PQ phase $\theta_{\phi}$ is well defined everywhere and is guiding the phases of the light quark condensates. 
   
The  reason why the question is non-trivial for the  $\eta'$ domain wall is that through such a wall $\theta_{\rm eff}$ changes by $2\pi$.  Correspondingly, QCD probes phases for which the  backreaction to the order parameter may become significant.
In particular, the corrections from beyond the dilute instanton gas approximation can become important. Notice that the analogous question does not arise for domain walls with $\pi^0$ winding.
 
Of course, an immediate technical obstacle for resolving  the internal structure of the pure-QCD  $\eta'$ domain wall is that within the wall the effective theory of the order parameter breaks down. However, we are not interested in an exact solution but only in an assurance  that the condensate is non-zero across the wall. 
   
To answer this question, a very useful guideline is provided by domain walls in ${\mathcal N}=1$ supersymmetric QCD (SQCD) \cite{Dvali:1996xe}. These walls are the closest cousins to the $\eta'$ domain walls discussed above.

The closest relative to QCD is an ${\mathcal N}=1$ supersymmetrized version of $SU(N)$ Yang-Mills (SYM). The role of the quarks in this theory is played by the fermionic partners of the gauge bosons, the gauginos, $\lambda$. Unlike the quarks in QCD, which transform in the fundamental representation, the gauginos transform in the adjoint representation of $SU(N)$. However, we expect that this difference is not essential for our discussion.

The theory possess an anomalous chiral symmetry, the $U(1)_R$-symmetry, 
\begin{equation} \label{RRR}  
    \lambda \rightarrow {\rm e}^{i\alpha} \lambda\,.   
\end{equation}
Since the gauginos transform in the adjoint representation of the $SU(N)$ gauge group, the $Z_N$ subgroup of 
$U(1)_R$ is anomaly free. 

It is well known that similarly to ordinary QCD, $SU(N)$ SYM theory confines and breaks the anomalous $U(1)_R$ symmetry through the gaugino condensate $\langle \bar{\lambda}{\lambda} \rangle \neq 0$.  The 't Hooft determinant generated by instantons represents  a $2N$-gaugino vertex. %not 2(N^2-1) vertex?
 
As pointed out in \cite{Dvali:1996xe}, this breaking gives rise to domain walls since the condensate spontaneously breaks the anomaly free $Z_N$-symmetry  
\begin{equation}
    \langle \bar{\lambda}{\lambda} \rangle
    \rightarrow {\rm e}^{i\frac{2\pi}{N}}  \langle \bar{\lambda}{\lambda}  \rangle \,.
\end{equation} 
Moreover, due to BPS properties of the wall, the wall tension can be computed exactly in terms of the gaugino condensate. 

It was conjectured in \cite{Witten:1997ep} that for large $N$ the wall acts as a sort of a $D$-brane for large-$N$ QCD string theory. This was further supported by the analysis of \cite{Dvali:1998ms, Dvali:1999pk} and of many subsequent papers. 

Now, while across the elementary wall the phase of the condensate changes by $2\pi/N$, the effective phase of the 't Hooft determinant changes by $2\pi$. Correspondingly, across the wall $\theta_{\rm eff}$ changes by $2\pi$ similarly to the $\eta'$ domain walls in ordinary QCD. Due to this, the question of validity of the phase degree of freedom within the two cases should be similar. 

For our purposes, the solution derived in \cite{Dvali:1998ms} is most informative. In this paper, the functional dependence of the gaugino condensate across the domain wall was evaluated explicitly. The solution is exact in large $N$. In this solution the gaugino condensate is non-zero across the wall and the phase degree of freedom (the analog of $\eta'$) is well-defined everywhere. Extrapolating this result to our case of ordinary QCD with quarks in the fundamental representation, we get the assurance that $\eta'$ remains well-defined across the wall. 

We thus conclude that pure QCD can support at least two types of string-wall systems, determined by the winding of $\pi^0$ and $\eta'$ respectively. 

\section{Cosmology} 
 
As we have seen, topologically non-trivial $\eta'$ or $\pi^0$ winding configurations of the QCD quark condensate will in general accompany axionic strings. Moreover, $\eta'$ or $\pi^0$ string-wall defects exist in pure QCD without involving a hidden axion. This allows for an entirely different view of $\theta$-vacuum cosmology that, to our knowledge, has 
not been considered previously. 

Strings can only have a cosmological relevance if the phase transition in which they are formed takes place after (or towards the end of) inflation. In most inflationary scenarios the reheating temperature  after inflation is larger than the present day QCD scale. This separation of present-day scales usually leads to the assumption that  QCD effects play no role in the early cosmology of the $\theta$-vacuum. In particular, the standard picture is based on two assumptions: \\
   
{\it 1)}   Early string cosmology is dominated by PQ strings of the hidden axion.\\
   
{\it 2)}   At the moment of string-formation the axion potential vanishes. Therefore, strings are free of domain walls.\\
  
We argue that the above picture is only valid in a restricted parameter space within a limited class of scenarios, whereas more generic scenarios are drastically different. In particular, the defects that dominate the early cosmology of the $\theta$-vacuum can be the string-wall systems produced by the QCD condensate, corresponding to $\eta'$ or pion windings. Due to this, in our discussion we will commonly refer to early defects as $\theta$-defects, rather than specifically as PQ strings.  
   
In order to explain the shortcomings of the standard scenario, let us note that the above assumptions are based on the large hierarchy of the present day scales 
\begin{equation} \label{PDscales}
    \frac{f_{\phi}}{\Lambda} \Big|_{\rm today} \gg 1 \,.  
\end{equation} 
However, the reasoning that maps this hierarchy onto the sequence of cosmological phase transitions contains a logical gap, implicitly assuming that the present day hierarchy of scales (\ref{PDscales}) can be extrapolated to the post-inflationary epoch. This assumption is not supported by any well-established understanding of the early cosmology and has already been challenged in \cite{Dvali:1995ce}. In order to understand this, let us focus on an immediate post-inflationary scenario of defect formation.

First, notice that a post-inflationary formation of $\theta$-defects does not necessarily correspond directly with the reheating temperature, as the transition could have taken place non-thermally, through direct interactions with the inflaton field, which we denote by $\Sigma$. 
   
In general, within an effective field theory, the potential of the PQ field is expected to depend on couplings to the inflaton (or other fields) of the type 
\begin{equation} \label{InfPQ}
    {\mathcal U} \left (\frac{\Sigma}{M} \right ) \Phi^*\Phi \,,
\end{equation} 
where $ {\mathcal U}$ is a generic function of $\Sigma$ and $M$ is a scale. Such couplings can change the VEV of the PQ field in a $\Sigma$-dependent way, thereby influencing the time of the phase transition as well as the tension of the PQ strings.  

However, the most important thing for the present discussion is the value of the QCD scale during the spontaneous breaking of the axial $U(1)_A$-symmetry. As shown in \cite{Dvali:1995ce}, the QCD gauge coupling could have become strong during or towards the end of inflation, resulting in a potential for $\theta_{\rm eff}$. 
As pointed out there, the gauge coupling can strongly depend on the inflaton or other fields (which themselves can strongly depend on the inflaton). This can cause the effective coupling strengths to vary significantly throughout the (post)inflationary epoch, either due to large excursions of the expectation value of the inflaton  (and its subsequent decay) or through fields whose VEVs depends on the inflaton (see e.g. \eqref{InfPQ}). Even if such a field-dependence is not taken into account by tree-level renormalizable operators (which has no a priory justification), it is expected to be generated both by loops as well as by high dimensional operators (for early discussions see \cite{Dvali:1995fb}).  
 
This situation is generic in inflationary scenarios emerging from string theory, such as e.g., brane inflation \cite{Dvali:1998pa, Dvali:2001fw}. This is, because in string theory, gauge couplings are dependent on fields, such as the dilaton. These fields, as argued in \cite{Dvali:1995ce}, become easily displaced during the inflationary epoch, due to their interactions with the inflaton. One important consequence of this displacement is a substantial change of the gauge couplings as compared to their present day values. 

However, it is not necessary to make any assumptions about the UV structure of the theory: Due to the EFT framework, it is natural to expect the appearance of operators like
\begin{equation} \label{Infgauge}
    \mathcal{W}\left (\frac{\Sigma}{M} \right ) {\rm tr} G_{\mu\nu}G^{\mu\nu}  \,, 
\end{equation} 
where $\mathcal{W}$ is a generic function of $\Sigma$ and $M$ is a scale. These operators effectively set the value of the gauge coupling through the relation
\begin{equation} \label{QCDgaugeInf}
    \frac{1}{g^2} = \mathcal{W}\left (\frac{\Sigma}{M} \right )\,.
\end{equation}   

Due to this, it is reasonable to expect that the gauge coupling depends on the inflaton, either directly or through other fields. 

Furthermore, it is natural to assume that its value during inflation differed significantly from its current value. 
%Hence, it is not justified to assume that the value of the QCD coupling today and during the inflationary epoch were the same. 
%Rather, it would be more conservative to take into account a large potential difference between the two values.  
      
Of course, whether at the moment of  symmetry breaking and string formation the gauge coupling was shifted towards stronger or weaker values, is scenario-dependent. However, since both the variety as well as the parameter space of inflationary models is large, the likelihood of actualization of the strong early epoch QCD scenario of \cite{Dvali:1995ce} appears rather high.

Thus, it is possible that the early epoch hierarchy between $f_\phi$ and $\Lambda$ is drastically different from its present day value \eqref{PDscales}. In fact, the hierarchy could be as extreme as
\begin{equation} \label{INscales}
    \frac{f_{\phi}}{\Lambda} \Big|_{\rm early} \ll 1 \,.  
\end{equation} 
In this case, not only the potential of $\theta_{\rm eff}$ is generated in the early epoch, but QCD dynamics fully dominates the cosmology of $\theta$-defects\footnote{This has immediate implication of lifting out the previous cosmological bound on the axion scale, $f_\phi > 10^{12} \ \rm GeV$ \cite{Preskill:1982cy, Dine:1982ah, Abbott:1982af}, since in the scenario of \cite{Dvali:1995ce}, the axion's coherent oscillations start way before the ordinary QCD temperatures. Correspondingly, even for a maximal initial amplitude, the axion energy gets efficiently red-shifted in the early epoch. 
For more recent implementations of this scenario, see \cite{Koutsangelas:2022lte} and references therein.}.
     
In particular, there are the following important implications. First, the axial $U(1)_A$ (or $U(1)_{PQ}$) cosmic strings, whether formed by the PQ or the QCD condensate, are immediately accompanied by the domain walls. More importantly, the string dynamics could have been dominated not by the PQ hidden axion but entirely by the phase of the QCD condensate. In fact, the hierarchy \eqref{INscales} could have been so extreme that, at the moment of quark condensation, the PQ field could could have been in the symmetric vacuum, $\Phi = 0$. This would imply that the main driving force of the  string-wall dynamics is not the PQ axion, but rather the quark composites, such as $\eta'$ or  $\pi^0$.

\subsection{Pion strings} 

While the $\eta'$-strings would always be accompanied by  domain walls, the story with pion strings is more subtle and deserves special attention. Their precise nature depends on the VEV of the Higgs field at the time of quark condensation. Since in this discussion we are not interested in a hidden axion, we restrict it to a pure SM with a single neutral Higgs denoted by 
$H \, = \, \rho(x) {\rm e}^{i\chi(x)}$\footnote{The other component of the Higgs doublet plays no role for our considerations, see also section VI.}. In order to clarify various regimes, let us consider the Yukawa couplings, restricting for simplicity, to a single generation of quarks
\begin{equation} \label{YHud}
    g_u H \bar{u}_Lu_R  \, + \, g_d H^* \bar{d}_Ld_R \, + 
    {\rm h.c.} \,.
\end{equation} 
The only exact chiral symmetry respected by these couplings is the local symmetry gauged by the $Z$-boson (\ref{gaugeZ}). Under it, the phases of the Higgs and the quark condensates transform as 
\begin{equation} %\label{gaugePhases1}
    \chi  \rightarrow   \chi   +  \beta\,, ~
    \theta_{u}  \rightarrow   \theta_{u}  -  \beta\,, ~  
    \theta_{d}  \rightarrow   \theta_{d}  +  \beta \,.
\end{equation} 
In the ordinary (present-day) SM vacuum, the VEV of the Higgs, $v$, is much larger than the quark condensate. Correspondingly, the combination of the phases that becomes the longitudinal component of the $Z$-boson, 
\begin{equation} %\label{gaugePhases1}
    v^2 \chi  -  f_u^2  \theta_{u}  + f_d^2 \theta_{d} \,, 
\end{equation} 
consists predominantly of the Higgs phase, with a tiny admixture from the phases of the quark condensate. 
The orthogonal degree of freedom corresponds to $\pi^0$, which gets its mass from the above Yukawa couplings. 

Now, in the early universe the story can be very different. In particular, if in the epoch of quark condensation, the Higgs boson had a smaller VEV, the Goldstone boson eaten up by the $Z$-boson would be $\pi^0$, with a small admixture from $\chi$\footnote{This admixture 
would be non-zero since even if the mass square of the Higgs, $m_H^2$, is large and positive, the quark condensate would nevertheless source a small VEV of the Higgs $v \sim \Lambda^3/m_H^2$.}. The cosmic strings formed in this phase transition would be local strings carrying the magnetic flux of the $Z$-boson
     \begin{equation} 
    \mathrm{flux} = 
    \frac{1}{g_z} \frac{v^2  n_{\chi}  - 
    f_{u}^2 \,  n_{\theta_{u}}  +  f_{d}^2  n_{\theta_{d}}}{
    v^2  +  
    f_{u}^2 \,  +  f_{d}^2 } \,,
\end{equation}
Since all the phases wind by the gauge transformation of the $Z$-boson, the flux is integer. For example, a minimal string with winding numbers $n_{\theta_{u}} = - n_{\theta_{d}} = - n_{\chi}= \pm 1$ carries a unit magnetic $\mathrm{flux} = \mp 1/g_z$. 
  
In light of the integer $Z$-flux, such pion strings share similarities 
with semi-local electroweak strings 
\cite{Achucarro:1999it}. However, the crucial difference is that the pion strings 
are formed by the winding of the phases 
of the QCD quark condensate. 
 
If, at the moment of quark condensation, we have $v \gg f_u,f_d$, the   story is opposite. In this case,  the low energy theory of $\pi^0$ has an approximate global symmetry, which is not touching 
the Higgs phase $\chi$. Under this symmetry global strings can form. For example, around such a string the winding numbers can be 
$n_{\theta_{u}} = - n_{\theta_{d}} = \pm 1$
and $n_{\chi} = 0$.  The magnetic flux of the $Z$-boson carried by such strings will be negligible. 
These strings will be the boundaries of $2\pi$ domain walls, formed due to explicit breaking of symmetry by the quark masses.  

A priori, both scenarios are equally plausible. This is due to the fact that, while the QCD gauge coupling can become strong through interactions of the type described by equation \eqref{Infgauge}, the inflaton can also dynamically alter the VEV of the Higgs field by coupling to it via \eqref{InfPQ}. Therefore, it is entirely model-dependent which of the two scenarios will play out.

The cosmology of such string-wall systems formed by $\pi^0$ or $\eta'$ in an epoch of early strong QCD \cite{Dvali:1995ce}, would be very different from the standard one. In particular, they would have a much higher wall tension, as the value of the condensate would necessarily be much larger. This leads to a much faster collapse, resulting in intense radiation, including in gravitational waves.  

To summarize, the new key point we would like to bring with this discussion is that, instead of the axion, $\eta'$ and pion string-walls can be the main driving force of the early cosmology of the $\theta$-vacuum. 

Notice, that since the early QCD scale can be much larger than the explicit breaking of $U(1)_A$ due to Yukawa couplings, other, heavier mesons, such as $\eta$ could be equally (or more) important for the winding. 
In fact, in scenarios with $\Lambda / v|_{early} \gg 1$ the number of flavors could even go beyond $N_f =3$, as now even the heavy quarks (those with $m_Q\gtrsim 1 \ \rm Gev$) might condense during the phase transition. Thus, there might be a much higher number of mesons involved than in the minimal picture\footnote{Using KSVZ type models, this would potentially even allow for $N_f \geqslant 6$.}. 

\subsection{Implications for QCD phase transition}
     
The string-wall systems produced by the QCD chiral condensate can have implications for the ordinary, thermal QCD phase transition. We assume that by this epoch the inflaton and other heavy VEVs are settled to their vacua, and the formation of the QCD condensate takes place according to the standard picture. Even in this case, a topologically non-trivial winding of the quark condensate phases can affect the story.    
      
As there is no significant hierarchy of scales, the strings and walls form essentially simultaneously below the QCD temperatures. That is, upon the formation of the quark condensate, its phase already possesses a well-defined minimum. Therefore, below the critical temperature, the absolute value  of the order parameter starts growing and at the same time its phase finds the minimum.      
           
However, the phases $\theta_u, \theta_d$ are ill-defined before the start of the transition. After the onset of the transition they are defined modulo $2\pi$ and each phase can relax towards the minimum from a different initial value, which they obtain at the start of the transition. This allows for the formation of domain walls, across which some of the phases change by $2\pi$.  
       
Understanding the characteristic size of the wall requires a more precise knowledge of the phase transition. Applying the intuition coming from the analysis of domain walls in weakly-coupled scalar theories \cite{Vilenkin:1982ks, Press:1989yh, Leite:2011sc}, their size can reach the Hubble scale.  Beyond this scale, it should be expected that these walls are bounded by strings, as the Hubble scale sets the upper limit on the correlation length.  
   
Due to the gradual Hubble expansion, soon after their formation, those strings will enter the horizon and cause the walls to collapse. Their energy is released in hadrons as well as in gravitational waves.   
 
 \section{Implications for heavy ion colliders?}
   
From the point of view of the low energy theory of mesons, the string-wall-systems formed by the QCD condensate represent solitonic states. An important question is the possibility of observing them in high energy collisions. In general, the formation of solitons or other non-perturbative states during the collision of a small number of high energy quanta is exponentially suppressed \cite{Brown:1992ay, Voloshin:1992rr, Argyres:1992np, Gorsky:1993ix, Libanov:1994ug, Libanov:1995gh, Son:1995wz, Dvali:2014ila, 
Addazi:2016ksu, 
Dvali:2018xoc, Monin:2018cbi, Dvali:2020wqi, 
Dvali:2022vzz}. 
    
This suppression can be understood by thinking of such processes in terms of effective $2 \rightarrow N$ $S$-matrix transitions. 
In a weakly interacting theory, a soliton of some characteristic size $L$ and mass $M \gg 1/L$ can be thought of as a coherent state. The constituent quanta of the state have characteristic energy $E \sim 1/L$ and occupation number $N \sim  (ML) \gg 1$. Hence, the formation of a soliton in a high-energy two-particle scattering process represents a transition process from two hard quanta into $N$ soft quanta (of energy $\sim 1/L$). On general grounds, it can be proven \cite{Dvali:2020wqi} that for a well-defined transition such a matrix element is bounded from above by a factor ${\rm e}^{-N}$. 

Overcoming this suppression is possible, but only if the final state soliton has a maximal microstate degeneracy. For $N$-particle coherent bound states, the maximal possible degeneracy scales as  $\sim {\rm e}^{N}$
\cite{Dvali:2020wqi}. Correspondingly, such states have maximal microstate entropy $S \sim N$. There exists evidence \cite{Dvali:2021ooc} that such a compensation takes place 
for certain multi-gluon states called the "color glass condensate" \cite{Gelis:2010nm}. However, for the string-wall solitons discussed in the present work, the degeneracy factor requires further investigation.

Given our present understanding, promising places to look for the experimental manifestations of these objects can be heavy ion collisions, such as at the LHC \cite{ATLAS:2008xda, *ALICE:2008ngc} and at the RIHC \cite{Harrison:2003sb, *Ozaki:1990sr}. Of course, at best, these collisions could produce small defects, not much exceeding the QCD length. However, any chance of catching glimpses of string-wall resonances would be of extraordinary importance.  \\

\section{Comment on the gauge axion} 

In \cite{Dvali:2005an} an alternative formulation of the axion was introduced, not relying on any anomalous global symmetry. Instead, in this formulation the axion represents an intrinsic part of the QCD gauge redundancy. Hence the term "gauge axion". Under the QCD gauge transformation of the gluons, 
$A_{\mu} \rightarrow U(x)A_{\mu} U^{\dagger}(x) 
+  U \partial_{\mu} U^{\dagger}$ with $U(x) \equiv e^{-i\omega(x)^bT^b}$, the gauge axion shifts as 
     \begin{equation} \label{Bgauge}
B_{\mu\nu} \, \rightarrow \, B_{\mu\nu} \, + \, \frac{1}{f}\Omega_{\mu\nu} \,,
\end{equation} 
where $\Omega_{\mu\nu} = {\rm tr} A_{[\mu}\partial_{\nu]}\omega$ and $f$ is the  gauge axion scale. 
     
The gauge axion $B_{\mu\nu}$ enters the Lagrangian via a unique gauge invariant combination 
\begin{align}
    C_{\mu\nu\alpha} - f\partial_{[\mu}B_{\nu\alpha]}
\end{align}
where $C_{\mu\nu\alpha}  \equiv {\rm tr} \left(A_{[\mu}\partial_{\nu}A_{\alpha]} + \frac{2}{3}A_{[\mu}A_{\nu}A_{\alpha]}\right)$ is the Chern-Simons 3-form of QCD, which under the QCD gauge transformation shifts as 
\begin{align}
    C_{\mu\nu\alpha} \rightarrow C_{\mu\nu\alpha} + \partial_{[\mu}\Omega_{\nu\alpha]}.
\end{align}   
The  gauge redundancy of $B_{\mu\nu}$ makes $\bar{\theta}$ unphysical and solves the strong-$CP$ problem to all orders in the operator expansion \cite{Dvali:2005an,Dvali:2022fdv}.  
      
At low energies, the  gauge axion can be dualized into  a pseudo-scalar phase with an effective Lagrangian given by \eqref{EcosA}. Correspondingly, the structure of domain walls discussed in the present work goes through the gauge axion formulation. 
     
However, the gauge axion cannot be UV-completed into a PQ complex scalar \cite{Dvali:2005an, Sakhelashvili:2021eid,Dvali:2022fdv}. Due to this, the  cosmic string structure of the gauge axion is different and requires a separate discussion. This will be given in the upcoming work \cite{gaugeaxion}. 
       
Finally, we wish to comment that the structure of string-wall systems produced purely via the QCD condensate, without the involvement of a hidden axion, such as $\eta'$ or $\pi^0$ string-wall systems, are not  sensitive to the PQ versus gauge nature of the axion and remain valid regardless of its presence.

\section{Summary and outlook} 
    
In this paper we have explored the winding configurations of QCD  quark condensates. In particular, the $2\pi$ domain walls bounded by cosmic strings, formed by windings of $\eta'$ or $\pi^0$ mesons. If a hidden axion  exists in the theory, such windings in general accompany the axionic string-wall system, affecting the physics of fermionic zero modes and hence anomaly inflow and superconductivity properties. 
  
However, the QCD condensate string-walls can exist independently of the hidden axion and can even play the dominant role in early cosmology.  In particular, they can be a potential source of gravitational waves and electromagnetic radiation.
%as well as of intense electromagnetic radiation.

In the low energy theory of mesons, such objects appear as solitonic states.  We discussed prospects of their production in heavy ion collisions.  

The string-wall structures described in the present paper 
 are fully supported by numerical results, which shall be discussed
 in the future work \cite{numerics}.

Finally, we would like to comment on possible string-wall structures formed by fermion condensates in other 
gauge sectors of the Standard Model and gravity. 

In this respect, we wish to point out that string-wall defects emerging in a theory of neutrino mass \cite{Dvali:2016uhn}
 have been studied in  \cite{Dvali:2021uvk}. These defects are formed due to the spontaneous breaking of a non-abelian neutrino-flavor symmetry by the neutrino condensate  \cite{Dvali:2013cpa}. 

Furthermore, we would like to comment that, according to the arguments given in \cite{Dvali:2024dlb}, the gravitational Eguchi-Hanson instantons  \cite{Eguchi:1978xp, *Eguchi:1978gw}, imply the presence of the condensate in spin-$3/2$ \cite{Hawking:1978ghb, *Konishi:1988mb, *Konishi:1989em} (rather than in spin-$1/2$) fermions. 
This condensate breaks the anomalous supersymmetric $R$-symmetry spontaneously.
The defects emerging from this breaking are $R$-string-wall systems. 
Since $R$-charges are shared by the Standard Model particles, it is generic for $R$-strings to carry a
$Z$-boson flux \cite{Dvali:1994qf}.

On a separate point, it was argued recently \cite{Dvali:2024zpc}, that 
the electroweak sector of the  Standard Model is accompanied by a new pseudo-scalar  meson, $\eta_w$, that shifts under the anomalous  $B+L$-symmetry. This degree of freedom mixes with the phase of a $B+L$-violating quark and lepton condensate formed by electroweak instantons.  It would be natural to investigate the associated string-wall systems. 
\\

{\bf Acknowledgments}

It is a pleasure to thank Maximilian Bachmaier, Archil Kobakhidze and Otari Sakhelashvili for discussions. This work was supported in part by the Humboldt Foundation under Humboldt Professorship Award, by the European Research Council Gravities Horizon Grant AO number: 850 173-6, by the Deutsche Forschungsgemeinschaft (DFG, German Research Foundation) under Germany's Excellence Strategy - EXC-2111 - 390814868, and Germany's Excellence Strategy under Excellence Cluster Origins. \\
 
Disclaimer: Funded by the European Union. Views and opinions expressed are however those of the authors only and do not necessarily reflect those of the European Union or European Research Council. Neither the European Union nor the granting authority can be held responsible for them.

\appendix
\section*{Appendix}
\label{sec:appendix}

In the massive quark case, the broken $U(1)_V$ gives rise to strings attached to domain walls.
This can be see from the energy of the system given in Eq. \eqref{eq:EwithM}.

The usual axionic string solution $\theta_\phi = l/R$ has energy density $\sigma_R=2 \pi (f_\phi^2/R +R \Lambda^4)$.
The energy of a string wall system also winding in $\eta'$ has lower energy.

The ansatz in Eq. \eqref{eq:ansatzeta} and \eqref{eq:ansatzphi} approximate a string wall system for $\eta'$ and $\phi$.
We perform the integration in the energy functional from Eq. \eqref{eq:EwithM} to find
\begin{align}  %\label{EwithM}
    \sigma (L_{\phi}, L_{\eta})\,  &= \, 2\pi f_{\phi}^2 \frac{1}{L_{\phi}}
    + 2\pi f_{\eta}^2 \frac{1}{L_{\eta}}  + 2 \pi \Lambda_m^4\, L_{\eta} +\\
    &\nonumber+ 2 \pi \Lambda^4 L_{\phi} \left ( 1 + \frac{L_{\phi}}{2\pi(L_{\phi} - L_{\eta})} \sin \left ( 2\pi \frac{L_{\eta}}{L_{\phi}} \right ) \right).
\end{align} 
To get an idea about the energy balance, we compare two regimes.  
  
First let us take $L_{\phi} = L_{\eta} \equiv  L$.   
In this regime the $\eta'$ and $\phi$ wind exactly opposite.
The minimization of energy with respect to $L$ gives the optimal winding length: 
\begin{equation}
    L  = \frac{f}{\Lambda_m^2},
\end{equation}  
where we used $f \equiv \sqrt{f_\phi^2 + f_{\eta}^2}$.
The energy of this solution is
\begin{equation} %\label{E1}
   \sigma_1  =  4\pi f \Lambda_m^2 \simeq  4\pi f_{\phi}  \Lambda_m^2\,.
\end{equation}  
The second regime is $L_{\eta} \ll L_{\phi}$, which approximately (i.e., in leading order in $L_{\eta} / L_{\phi}$ expansion) gives: 
\begin{align}  %\label{EwithM}
    \sigma (L_{\phi}, L_{\eta})
    &\simeq \, 2\pi f_{\phi}^2 \frac{1}{L_{\phi}} +
    + 2\pi f_{\eta}^2 \frac{1}{L_{\eta}}  + \\
    \nonumber &+2\pi\Lambda_m^4\, L_{\eta} + 2\pi \Lambda^4 L_{\phi} \left ( 1 + \frac{L_{\eta}}{L_{\phi}} \right).
\end{align} 
Extremizing the energy gives
\begin{equation} %\label{L2}
   L_{\phi} = \frac{f_{\phi}}{\Lambda^2}\,, \quad  L_{\eta} = \frac{f_{\eta}}{\sqrt{\Lambda_m^4+\Lambda^4}},
\end{equation}    
with corresponding energy
\begin{equation} %\label{E2}
   \sigma_2  =  4\pi f_{\phi} \Lambda^2 +  4\pi f_{\eta} \sqrt{\Lambda_m^4+\Lambda^4}\,.
\end{equation} 
Taking into account that $\Lambda \gg \Lambda_m$, we have
\begin{equation} %\label{E2}
   \frac{\sigma_2}{\sigma_1} \simeq \frac{\Lambda^2}{\Lambda^2_m}\gg 1\,, 
\end{equation} 
from which we conclude that the regime $L_{\phi} = L_{\eta} \equiv  L$ is energetically more favorable.

\bibliographystyle{utphys}
\bibliography{etawalls.bib}

\providecommand{\href}[2]{#2}\begingroup\raggedright\begin{thebibliography}{10}

\bibitem{Witten:1984eb}
E.~Witten, ``{Superconducting Strings},'' \href{http://dx.doi.org/10.1016/0550-3213(85)90022-7}{{\em Nucl. Phys. B} {\bfseries 249} (1985) 557--592}.

\bibitem{Callan:1984sa}
C.~G. Callan, Jr. and J.~A. Harvey, ``{Anomalies and Fermion Zero Modes on Strings and Domain Walls},'' \href{http://dx.doi.org/10.1016/0550-3213(85)90489-4}{{\em Nucl. Phys. B} {\bfseries 250} (1985) 427--436}.

\bibitem{Dvali:2005an}
G.~Dvali, ``{Three-form gauging of axion symmetries and gravity},'' \href{http://arxiv.org/abs/hep-th/0507215}{{\ttfamily arXiv:hep-th/0507215}}.

\bibitem{Dvali:1995ce}
G.~R. Dvali, ``{Removing the cosmological bound on the axion scale},'' \href{http://arxiv.org/abs/hep-ph/9505253}{{\ttfamily arXiv:hep-ph/9505253}}.

\bibitem{Callan:1976je}
C.~G. Callan, Jr., R.~F. Dashen, and D.~J. Gross, ``{The Structure of the Gauge Theory Vacuum},'' \href{http://dx.doi.org/10.1016/0370-2693(76)90277-X}{{\em Phys. Lett. B} {\bfseries 63} (1976) 334--340}.

\bibitem{Jackiw:1976pf}
R.~Jackiw and C.~Rebbi, ``{Vacuum Periodicity in a Yang-Mills Quantum Theory},'' \href{http://dx.doi.org/10.1103/PhysRevLett.37.172}{{\em Phys. Rev. Lett.} {\bfseries 37} (1976) 172--175}.

\bibitem{Baluni:1978rf}
V.~Baluni, ``{CP Violating Effects in QCD},'' \href{http://dx.doi.org/10.1103/PhysRevD.19.2227}{{\em Phys. Rev. D} {\bfseries 19} (1979) 2227--2230}.

\bibitem{Crewther:1979pi}
R.~J. Crewther, P.~Di~Vecchia, G.~Veneziano, and E.~Witten, ``{Chiral Estimate of the Electric Dipole Moment of the Neutron in Quantum Chromodynamics},'' \href{http://dx.doi.org/10.1016/0370-2693(79)90128-X}{{\em Phys. Lett. B} {\bfseries 88} (1979) 123}. [Erratum: Phys.Lett.B 91, 487 (1980)].

\bibitem{Baker:2006ts}
C.~A. Baker {\em et~al.}, ``{An Improved experimental limit on the electric dipole moment of the neutron},'' \href{http://dx.doi.org/10.1103/PhysRevLett.97.131801}{{\em Phys. Rev. Lett.} {\bfseries 97} (2006) 131801}, \href{http://arxiv.org/abs/hep-ex/0602020}{{\ttfamily arXiv:hep-ex/0602020}}.

\bibitem{Pendlebury:2015lrz}
J.~M. Pendlebury {\em et~al.}, ``{Revised experimental upper limit on the electric dipole moment of the neutron},'' \href{http://dx.doi.org/10.1103/PhysRevD.92.092003}{{\em Phys. Rev. D} {\bfseries 92} no.~9, (2015) 092003}, \href{http://arxiv.org/abs/1509.04411}{{\ttfamily arXiv:1509.04411 [hep-ex]}}.

\bibitem{Graner:2016ses}
B.~Graner, Y.~Chen, E.~G. Lindahl, and B.~R. Heckel, ``{Reduced Limit on the Permanent Electric Dipole Moment of Hg199},'' \href{http://dx.doi.org/10.1103/PhysRevLett.116.161601}{{\em Phys. Rev. Lett.} {\bfseries 116} no.~16, (2016) 161601}, \href{http://arxiv.org/abs/1601.04339}{{\ttfamily arXiv:1601.04339 [physics.atom-ph]}}. [Erratum: Phys.Rev.Lett. 119, 119901 (2017)].

\bibitem{Ellis:1976fn}
J.~R. Ellis, M.~K. Gaillard, and D.~V. Nanopoulos, ``{Lefthanded Currents and CP Violation},'' \href{http://dx.doi.org/10.1016/0550-3213(76)90203-0}{{\em Nucl. Phys. B} {\bfseries 109} (1976) 213--243}.

\bibitem{Shabalin:1978rs}
E.~P. Shabalin, ``{Electric Dipole Moment of Quark in a Gauge Theory with Left-Handed Currents},'' {\em Sov. J. Nucl. Phys.} {\bfseries 28} (1978) 75.

\bibitem{Ellis:1978hq}
J.~R. Ellis and M.~K. Gaillard, ``{Strong and Weak CP Violation},'' \href{http://dx.doi.org/10.1016/0550-3213(79)90297-9}{{\em Nucl. Phys. B} {\bfseries 150} (1979) 141--162}.

\bibitem{Peccei:1977hh}
R.~D. Peccei and H.~R. Quinn, ``{CP Conservation in the Presence of Instantons},'' \href{http://dx.doi.org/10.1103/PhysRevLett.38.1440}{{\em Phys. Rev. Lett.} {\bfseries 38} (1977) 1440--1443}.

\bibitem{Peccei:1977ur}
R.~D. Peccei and H.~R. Quinn, ``{Constraints Imposed by CP Conservation in the Presence of Instantons},'' \href{http://dx.doi.org/10.1103/PhysRevD.16.1791}{{\em Phys. Rev. D} {\bfseries 16} (1977) 1791--1797}.

\bibitem{Weinberg:1977ma}
S.~Weinberg, ``{A New Light Boson?},'' \href{http://dx.doi.org/10.1103/PhysRevLett.40.223}{{\em Phys. Rev. Lett.} {\bfseries 40} (1978) 223--226}.

\bibitem{Wilczek:1977pj}
F.~Wilczek, ``{Problem of Strong $P$ and $T$ Invariance in the Presence of Instantons},'' \href{http://dx.doi.org/10.1103/PhysRevLett.40.279}{{\em Phys. Rev. Lett.} {\bfseries 40} (1978) 279--282}.

\bibitem{Vafa:1984xg}
C.~Vafa and E.~Witten, ``{Parity Conservation in QCD},'' \href{http://dx.doi.org/10.1103/PhysRevLett.53.535}{{\em Phys. Rev. Lett.} {\bfseries 53} (1984) 535}.

\bibitem{tHooft:1976rip}
G.~'t~Hooft, ``{Symmetry Breaking Through Bell-Jackiw Anomalies},'' \href{http://dx.doi.org/10.1103/PhysRevLett.37.8}{{\em Phys. Rev. Lett.} {\bfseries 37} (1976) 8--11}.

\bibitem{tHooft:1976snw}
G.~'t~Hooft, ``{Computation of the Quantum Effects Due to a Four-Dimensional Pseudoparticle},'' \href{http://dx.doi.org/10.1103/PhysRevD.14.3432}{{\em Phys. Rev. D} {\bfseries 14} (1976) 3432--3450}. [Erratum: Phys.Rev.D 18, 2199 (1978)].

\bibitem{Dvali:2005ws}
G.~Dvali, R.~Jackiw, and S.-Y. Pi, ``{Topological mass generation in four dimensions},'' \href{http://dx.doi.org/10.1103/PhysRevLett.96.081602}{{\em Phys. Rev. Lett.} {\bfseries 96} (2006) 081602}, \href{http://arxiv.org/abs/hep-th/0511175}{{\ttfamily arXiv:hep-th/0511175}}.

\bibitem{Dvali:2013cpa}
G.~Dvali, S.~Folkerts, and A.~Franca, ``{How neutrino protects the axion},'' \href{http://dx.doi.org/10.1103/PhysRevD.89.105025}{{\em Phys. Rev. D} {\bfseries 89} no.~10, (2014) 105025}, \href{http://arxiv.org/abs/1312.7273}{{\ttfamily arXiv:1312.7273 [hep-th]}}.

\bibitem{FlavourLatticeAveragingGroupFLAG:2024oxs}
{\bfseries Flavour Lattice Averaging Group (FLAG)} Collaboration, Y.~Aoki {\em et~al.}, ``{FLAG Review 2024},'' \href{http://arxiv.org/abs/2411.04268}{{\ttfamily arXiv:2411.04268 [hep-lat]}}.

\bibitem{numerics}
M.~Bachmaier, G.~Dvali, L.~Komisel, and A.~Stuhlfauth in progress.

\bibitem{Jackiw:1981ee}
R.~Jackiw and P.~Rossi, ``{Zero Modes of the Vortex - Fermion System},'' \href{http://dx.doi.org/10.1016/0550-3213(81)90044-4}{{\em Nucl. Phys. B} {\bfseries 190} (1981) 681--691}.

\bibitem{Ostriker:1986xc}
J.~P. Ostriker, A.~C. Thompson, and E.~Witten, ``{Cosmological Effects of Superconducting Strings},'' \href{http://dx.doi.org/10.1016/0370-2693(86)90301-1}{{\em Phys. Lett. B} {\bfseries 180} (1986) 231--239}.

\bibitem{Lazarides:1984zq}
G.~Lazarides and Q.~Shafi, ``{Superconducting Strings in Axion Models},'' \href{http://dx.doi.org/10.1016/0370-2693(85)91398-X}{{\em Phys. Lett. B} {\bfseries 151} (1985) 123--126}.

\bibitem{Naculich:1987ci}
S.~G. Naculich, ``{Axionic Strings: Covariant Anomalies and Bosonization of Chiral Zero Modes},'' \href{http://dx.doi.org/10.1016/0550-3213(88)90400-2}{{\em Nucl. Phys. B} {\bfseries 296} (1988) 837--867}.

\bibitem{Harvey:2000yg}
J.~A. Harvey and O.~Ruchayskiy, ``{The Local structure of anomaly inflow},'' \href{http://dx.doi.org/10.1088/1126-6708/2001/06/044}{{\em JHEP} {\bfseries 06} (2001) 044}, \href{http://arxiv.org/abs/hep-th/0007037}{{\ttfamily arXiv:hep-th/0007037}}.

\bibitem{Agrawal:2019lkr}
P.~Agrawal, A.~Hook, and J.~Huang, ``{A CMB Millikan experiment with cosmic axiverse strings},'' \href{http://dx.doi.org/10.1007/JHEP07(2020)138}{{\em JHEP} {\bfseries 07} (2020) 138}, \href{http://arxiv.org/abs/1912.02823}{{\ttfamily arXiv:1912.02823 [astro-ph.CO]}}.

\bibitem{Agrawal:2020euj}
P.~Agrawal, A.~Hook, J.~Huang, and G.~Marques-Tavares, ``{Axion string signatures: a cosmological plasma collider},'' \href{http://dx.doi.org/10.1007/JHEP01(2022)103}{{\em JHEP} {\bfseries 01} (2022) 103}, \href{http://arxiv.org/abs/2010.15848}{{\ttfamily arXiv:2010.15848 [hep-ph]}}.

\bibitem{Fukuda:2020kym}
H.~Fukuda, A.~V. Manohar, H.~Murayama, and O.~Telem, ``{Axion strings are superconducting},'' \href{http://dx.doi.org/10.1007/JHEP06(2021)052}{{\em JHEP} {\bfseries 06} (2021) 052}, \href{http://arxiv.org/abs/2010.02763}{{\ttfamily arXiv:2010.02763 [hep-ph]}}.

\bibitem{Bagherian:2023jxy}
H.~Bagherian, K.~Fraser, S.~Homiller, and J.~Stout, ``{Zero modes of massive fermions delocalize from axion strings},'' \href{http://dx.doi.org/10.1007/JHEP05(2024)079}{{\em JHEP} {\bfseries 05} (2024) 079}, \href{http://arxiv.org/abs/2310.01476}{{\ttfamily arXiv:2310.01476 [hep-th]}}.

\bibitem{Kim:1979if}
J.~E. Kim, ``{Weak Interaction Singlet and Strong CP Invariance},'' \href{http://dx.doi.org/10.1103/PhysRevLett.43.103}{{\em Phys. Rev. Lett.} {\bfseries 43} (1979) 103}.

\bibitem{Shifman:1979if}
M.~A. Shifman, A.~I. Vainshtein, and V.~I. Zakharov, ``{Can Confinement Ensure Natural CP Invariance of Strong Interactions?},'' \href{http://dx.doi.org/10.1016/0550-3213(80)90209-6}{{\em Nucl. Phys. B} {\bfseries 166} (1980) 493--506}.

\bibitem{Dine:1981rt}
M.~Dine, W.~Fischler, and M.~Srednicki, ``{A Simple Solution to the Strong CP Problem with a Harmless Axion},'' \href{http://dx.doi.org/10.1016/0370-2693(81)90590-6}{{\em Phys. Lett. B} {\bfseries 104} (1981) 199--202}.

\bibitem{Zhitnitsky:1980tq}
A.~R. Zhitnitsky, ``{On Possible Suppression of the Axion Hadron Interactions. (In Russian)},'' {\em Sov. J. Nucl. Phys.} {\bfseries 31} (1980) 260.

\bibitem{Dvali:1993sg}
G.~R. Dvali and G.~Senjanovic, ``{Topologically stable electroweak flux tubes},'' \href{http://dx.doi.org/10.1103/PhysRevLett.71.2376}{{\em Phys. Rev. Lett.} {\bfseries 71} (1993) 2376--2379}, \href{http://arxiv.org/abs/hep-ph/9305278}{{\ttfamily arXiv:hep-ph/9305278}}.

\bibitem{Dvali:1996xe}
G.~R. Dvali and M.~A. Shifman, ``{Domain walls in strongly coupled theories},'' \href{http://dx.doi.org/10.1016/S0370-2693(97)00131-7}{{\em Phys. Lett. B} {\bfseries 396} (1997) 64--69}, \href{http://arxiv.org/abs/hep-th/9612128}{{\ttfamily arXiv:hep-th/9612128}}. [Erratum: Phys.Lett.B 407, 452 (1997)].

\bibitem{Witten:1997ep}
E.~Witten, ``{Branes and the dynamics of QCD},'' \href{http://dx.doi.org/10.1016/S0550-3213(97)00648-2}{{\em Nucl. Phys. B} {\bfseries 507} (1997) 658--690}, \href{http://arxiv.org/abs/hep-th/9706109}{{\ttfamily arXiv:hep-th/9706109}}.

\bibitem{Dvali:1998ms}
G.~R. Dvali and Z.~Kakushadze, ``{Large N domain walls as D-branes for N=1 QCD string},'' \href{http://dx.doi.org/10.1016/S0550-3213(98)00683-X}{{\em Nucl. Phys. B} {\bfseries 537} (1999) 297--316}, \href{http://arxiv.org/abs/hep-th/9807140}{{\ttfamily arXiv:hep-th/9807140}}.

\bibitem{Dvali:1999pk}
G.~R. Dvali, G.~Gabadadze, and Z.~Kakushadze, ``{BPS domain walls in large N supersymmetric QCD},'' \href{http://dx.doi.org/10.1016/S0550-3213(99)00562-3}{{\em Nucl. Phys. B} {\bfseries 562} (1999) 158--180}, \href{http://arxiv.org/abs/hep-th/9901032}{{\ttfamily arXiv:hep-th/9901032}}.

\bibitem{Dvali:1995fb}
G.~R. Dvali, ``{Inflation induced SUSY breaking and flat vacuum directions},'' \href{http://dx.doi.org/10.1016/0370-2693(95)00503-D}{{\em Phys. Lett. B} {\bfseries 355} (1995) 78--84}, \href{http://arxiv.org/abs/hep-ph/9503375}{{\ttfamily arXiv:hep-ph/9503375}}.

\bibitem{Dvali:1998pa}
G.~R. Dvali and S.~H.~H. Tye, ``{Brane inflation},'' \href{http://dx.doi.org/10.1016/S0370-2693(99)00132-X}{{\em Phys. Lett. B} {\bfseries 450} (1999) 72--82}, \href{http://arxiv.org/abs/hep-ph/9812483}{{\ttfamily arXiv:hep-ph/9812483}}.

\bibitem{Dvali:2001fw}
G.~R. Dvali, Q.~Shafi, and S.~Solganik, ``{D-brane inflation},'' in {\em {4th European Meeting From the Planck Scale to the Electroweak Scale}}.
\newblock 5, 2001.
\newblock \href{http://arxiv.org/abs/hep-th/0105203}{{\ttfamily arXiv:hep-th/0105203}}.

\bibitem{Preskill:1982cy}
J.~Preskill, M.~B. Wise, and F.~Wilczek, ``{Cosmology of the Invisible Axion},'' \href{http://dx.doi.org/10.1016/0370-2693(83)90637-8}{{\em Phys. Lett. B} {\bfseries 120} (1983) 127--132}.

\bibitem{Dine:1982ah}
M.~Dine and W.~Fischler, ``{The Not So Harmless Axion},'' \href{http://dx.doi.org/10.1016/0370-2693(83)90639-1}{{\em Phys. Lett. B} {\bfseries 120} (1983) 137--141}.

\bibitem{Abbott:1982af}
L.~F. Abbott and P.~Sikivie, ``{A Cosmological Bound on the Invisible Axion},'' \href{http://dx.doi.org/10.1016/0370-2693(83)90638-X}{{\em Phys. Lett. B} {\bfseries 120} (1983) 133--136}.

\bibitem{Koutsangelas:2022lte}
E.~Koutsangelas, ``{Removing the cosmological bound on the axion scale in the Kim-Shifman-Vainshtein-Zakharov and Dine-Fischler-Srednicki-Zhitnitsky models},'' \href{http://dx.doi.org/10.1103/PhysRevD.107.095009}{{\em Phys. Rev. D} {\bfseries 107} no.~9, (2023) 095009}, \href{http://arxiv.org/abs/2212.07822}{{\ttfamily arXiv:2212.07822 [hep-ph]}}.

\bibitem{Achucarro:1999it}
A.~Achucarro and T.~Vachaspati, ``{Semilocal and electroweak strings},'' \href{http://dx.doi.org/10.1016/S0370-1573(99)00103-9}{{\em Phys. Rept.} {\bfseries 327} (2000) 347--426}, \href{http://arxiv.org/abs/hep-ph/9904229}{{\ttfamily arXiv:hep-ph/9904229}}.

\bibitem{Vilenkin:1982ks}
A.~Vilenkin and A.~E. Everett, ``{Cosmic Strings and Domain Walls in Models with Goldstone and PseudoGoldstone Bosons},'' \href{http://dx.doi.org/10.1103/PhysRevLett.48.1867}{{\em Phys. Rev. Lett.} {\bfseries 48} (1982) 1867--1870}.

\bibitem{Press:1989yh}
W.~H. Press, B.~S. Ryden, and D.~N. Spergel, ``{Dynamical Evolution of Domain Walls in an Expanding Universe},'' \href{http://dx.doi.org/10.1086/168151}{{\em Astrophys. J.} {\bfseries 347} (1989) 590--604}.

\bibitem{Leite:2011sc}
A.~M.~M. Leite and C.~J. A.~P. Martins, ``{Scaling Properties of Domain Wall Networks},'' \href{http://dx.doi.org/10.1103/PhysRevD.84.103523}{{\em Phys. Rev. D} {\bfseries 84} (2011) 103523}, \href{http://arxiv.org/abs/1110.3486}{{\ttfamily arXiv:1110.3486 [hep-ph]}}.

\bibitem{Brown:1992ay}
L.~S. Brown, ``{Summing tree graphs at threshold},'' \href{http://dx.doi.org/10.1103/PhysRevD.46.R4125}{{\em Phys. Rev. D} {\bfseries 46} (1992) R4125--R4127}, \href{http://arxiv.org/abs/hep-ph/9209203}{{\ttfamily arXiv:hep-ph/9209203}}.

\bibitem{Voloshin:1992rr}
M.~B. Voloshin, ``{Estimate of the onset of nonperturbative particle production at high-energy in a scalar theory},'' \href{http://dx.doi.org/10.1016/0370-2693(92)90901-F}{{\em Phys. Lett. B} {\bfseries 293} (1992) 389--394}.

\bibitem{Argyres:1992np}
E.~N. Argyres, R.~H.~P. Kleiss, and C.~G. Papadopoulos, ``{Amplitude estimates for multi - Higgs production at high-energies},'' \href{http://dx.doi.org/10.1016/0550-3213(93)90140-K}{{\em Nucl. Phys. B} {\bfseries 391} (1993) 42--56}.

\bibitem{Gorsky:1993ix}
A.~S. Gorsky and M.~B. Voloshin, ``{Nonperturbative production of multiboson states and quantum bubbles},'' \href{http://dx.doi.org/10.1103/PhysRevD.48.3843}{{\em Phys. Rev. D} {\bfseries 48} (1993) 3843--3851}, \href{http://arxiv.org/abs/hep-ph/9305219}{{\ttfamily arXiv:hep-ph/9305219}}.

\bibitem{Libanov:1994ug}
M.~V. Libanov, V.~A. Rubakov, D.~T. Son, and S.~V. Troitsky, ``{Exponentiation of multiparticle amplitudes in scalar theories},'' \href{http://dx.doi.org/10.1103/PhysRevD.50.7553}{{\em Phys. Rev. D} {\bfseries 50} (1994) 7553--7569}, \href{http://arxiv.org/abs/hep-ph/9407381}{{\ttfamily arXiv:hep-ph/9407381}}.

\bibitem{Libanov:1995gh}
M.~V. Libanov, D.~T. Son, and S.~V. Troitsky, ``{Exponentiation of multiparticle amplitudes in scalar theories. 2. Universality of the exponent},'' \href{http://dx.doi.org/10.1103/PhysRevD.52.3679}{{\em Phys. Rev. D} {\bfseries 52} (1995) 3679--3687}, \href{http://arxiv.org/abs/hep-ph/9503412}{{\ttfamily arXiv:hep-ph/9503412}}.

\bibitem{Son:1995wz}
D.~T. Son, ``{Semiclassical approach for multiparticle production in scalar theories},'' \href{http://dx.doi.org/10.1016/0550-3213(96)00386-0}{{\em Nucl. Phys. B} {\bfseries 477} (1996) 378--406}, \href{http://arxiv.org/abs/hep-ph/9505338}{{\ttfamily arXiv:hep-ph/9505338}}.

\bibitem{Dvali:2014ila}
G.~Dvali, C.~Gomez, R.~S. Isermann, D.~L\"ust, and S.~Stieberger, ``{Black hole formation and classicalization in ultra-Planckian 2\textrightarrow{}N scattering},'' \href{http://dx.doi.org/10.1016/j.nuclphysb.2015.02.004}{{\em Nucl. Phys. B} {\bfseries 893} (2015) 187--235}, \href{http://arxiv.org/abs/1409.7405}{{\ttfamily arXiv:1409.7405 [hep-th]}}.

\bibitem{Addazi:2016ksu}
A.~Addazi, M.~Bianchi, and G.~Veneziano, ``{Glimpses of black hole formation/evaporation in highly inelastic, ultra-planckian string collisions},'' \href{http://dx.doi.org/10.1007/JHEP02(2017)111}{{\em JHEP} {\bfseries 02} (2017) 111}, \href{http://arxiv.org/abs/1611.03643}{{\ttfamily arXiv:1611.03643 [hep-th]}}.

\bibitem{Dvali:2018xoc}
G.~Dvali, ``{Classicalization Clearly: Quantum Transition into States of Maximal Memory Storage Capacity},'' \href{http://arxiv.org/abs/1804.06154}{{\ttfamily arXiv:1804.06154 [hep-th]}}.

\bibitem{Monin:2018cbi}
A.~Monin, ``{Inconsistencies of higgsplosion},'' \href{http://arxiv.org/abs/1808.05810}{{\ttfamily arXiv:1808.05810 [hep-th]}}.

\bibitem{Dvali:2020wqi}
G.~Dvali, ``{Entropy Bound and Unitarity of Scattering Amplitudes},'' \href{http://dx.doi.org/10.1007/JHEP03(2021)126}{{\em JHEP} {\bfseries 03} (2021) 126}, \href{http://arxiv.org/abs/2003.05546}{{\ttfamily arXiv:2003.05546 [hep-th]}}.

\bibitem{Dvali:2022vzz}
G.~Dvali and L.~Eisemann, ``{Perturbative understanding of nonperturbative processes and quantumization versus classicalization},'' \href{http://dx.doi.org/10.1103/PhysRevD.106.125019}{{\em Phys. Rev. D} {\bfseries 106} no.~12, (2022) 125019}, \href{http://arxiv.org/abs/2211.02618}{{\ttfamily arXiv:2211.02618 [hep-th]}}.

\bibitem{Dvali:2021ooc}
G.~Dvali and R.~Venugopalan, ``{Classicalization and unitarization of wee partons in QCD and gravity: The CGC-black hole correspondence},'' \href{http://dx.doi.org/10.1103/PhysRevD.105.056026}{{\em Phys. Rev. D} {\bfseries 105} no.~5, (2022) 056026}, \href{http://arxiv.org/abs/2106.11989}{{\ttfamily arXiv:2106.11989 [hep-th]}}.

\bibitem{Gelis:2010nm}
F.~Gelis, E.~Iancu, J.~Jalilian-Marian, and R.~Venugopalan, ``{The Color Glass Condensate},'' \href{http://dx.doi.org/10.1146/annurev.nucl.010909.083629}{{\em Ann. Rev. Nucl. Part. Sci.} {\bfseries 60} (2010) 463--489}, \href{http://arxiv.org/abs/1002.0333}{{\ttfamily arXiv:1002.0333 [hep-ph]}}.

\bibitem{ATLAS:2008xda}
{\bfseries ATLAS} Collaboration, G.~Aad {\em et~al.}, ``{The ATLAS Experiment at the CERN Large Hadron Collider},'' \href{http://dx.doi.org/10.1088/1748-0221/3/08/S08003}{{\em JINST} {\bfseries 3} (2008) S08003}.

\bibitem{ALICE:2008ngc}
{\bfseries ALICE} Collaboration, K.~Aamodt {\em et~al.}, ``{The ALICE experiment at the CERN LHC},'' \href{http://dx.doi.org/10.1088/1748-0221/3/08/S08002}{{\em JINST} {\bfseries 3} (2008) S08002}.

\bibitem{Harrison:2003sb}
M.~Harrison, T.~Ludlam, and S.~Ozaki, ``{RHIC project overview},'' \href{http://dx.doi.org/10.1016/S0168-9002(02)01937-X}{{\em Nucl. Instrum. Meth. A} {\bfseries 499} (2003) 235--244}.

\bibitem{Ozaki:1990sr}
S.~Ozaki, ``{The Relativistic heavy ion collider at Brookhaven},'' \href{http://dx.doi.org/10.1016/0375-9474(91)90320-6}{{\em Nucl. Phys. A} {\bfseries 525} (1991) 125C--132C}.

\bibitem{Dvali:2022fdv}
G.~Dvali, ``{Strong-$CP$ with and without gravity},'' \href{http://arxiv.org/abs/2209.14219}{{\ttfamily arXiv:2209.14219 [hep-ph]}}.

\bibitem{Sakhelashvili:2021eid}
O.~Sakhelashvili, ``{Consistency of the dual formulation of axion solutions to the strong CP problem},'' \href{http://dx.doi.org/10.1103/PhysRevD.105.085020}{{\em Phys. Rev. D} {\bfseries 105} no.~8, (2022) 085020}, \href{http://arxiv.org/abs/2110.03386}{{\ttfamily arXiv:2110.03386 [hep-th]}}.

\bibitem{gaugeaxion}
G.~Dvali, L.~Komisel, and A.~Stuhlfauth in progress.

\bibitem{Dvali:2016uhn}
G.~Dvali and L.~Funcke, ``{Small neutrino masses from gravitational \ensuremath{\theta}-term},'' \href{http://dx.doi.org/10.1103/PhysRevD.93.113002}{{\em Phys. Rev. D} {\bfseries 93} no.~11, (2016) 113002}, \href{http://arxiv.org/abs/1602.03191}{{\ttfamily arXiv:1602.03191 [hep-ph]}}.

\bibitem{Dvali:2021uvk}
G.~Dvali, L.~Funcke, and T.~Vachaspati, ``{Time- and Space-Varying Neutrino Mass Matrix from Soft Topological Defects},'' \href{http://dx.doi.org/10.1103/PhysRevLett.130.091601}{{\em Phys. Rev. Lett.} {\bfseries 130} no.~9, (2023) 091601}, \href{http://arxiv.org/abs/2112.02107}{{\ttfamily arXiv:2112.02107 [hep-ph]}}.

\bibitem{Dvali:2024dlb}
G.~Dvali, A.~Kobakhidze, and O.~Sakhelashvili, ``{Hint to supersymmetry from the GR vacuum},'' \href{http://dx.doi.org/10.1103/PhysRevD.110.086008}{{\em Phys. Rev. D} {\bfseries 110} no.~8, (2024) 086008}, \href{http://arxiv.org/abs/2406.18402}{{\ttfamily arXiv:2406.18402 [hep-th]}}.

\bibitem{Eguchi:1978xp}
T.~Eguchi and A.~J. Hanson, ``{Asymptotically Flat Selfdual Solutions to Euclidean Gravity},'' \href{http://dx.doi.org/10.1016/0370-2693(78)90566-X}{{\em Phys. Lett. B} {\bfseries 74} (1978) 249--251}.

\bibitem{Eguchi:1978gw}
T.~Eguchi and A.~J. Hanson, ``{Selfdual Solutions to Euclidean Gravity},'' \href{http://dx.doi.org/10.1016/0003-4916(79)90282-3}{{\em Annals Phys.} {\bfseries 120} (1979) 82}.

\bibitem{Hawking:1978ghb}
S.~W. Hawking and C.~N. Pope, ``{Symmetry Breaking by Instantons in Supergravity},'' \href{http://dx.doi.org/10.1016/0550-3213(78)90073-1}{{\em Nucl. Phys. B} {\bfseries 146} (1978) 381--392}.

\bibitem{Konishi:1988mb}
K.~Konishi, N.~Magnoli, and H.~Panagopoulos, ``{Spontaneous Breaking of Local Supersymmetry by Gravitational Instantons},'' \href{http://dx.doi.org/10.1016/0550-3213(88)90239-8}{{\em Nucl. Phys. B} {\bfseries 309} (1988) 201}.

\bibitem{Konishi:1989em}
K.~Konishi, N.~Magnoli, and H.~Panagopoulos, ``{Generation of Mass Hierarchies and Gravitational Instanton Induced Supersymmetry Breaking},'' \href{http://dx.doi.org/10.1016/0550-3213(89)90151-X}{{\em Nucl. Phys. B} {\bfseries 323} (1989) 441--458}.

\bibitem{Dvali:1994qf}
G.~R. Dvali and G.~Senjanovic, ``{Topologically stable Z strings in the supersymmetric Standard Model},'' \href{http://dx.doi.org/10.1016/0370-2693(94)90943-1}{{\em Phys. Lett. B} {\bfseries 331} (1994) 63--68}, \href{http://arxiv.org/abs/hep-ph/9403277}{{\ttfamily arXiv:hep-ph/9403277}}.

\bibitem{Dvali:2024zpc}
G.~Dvali, A.~Kobakhidze, and O.~Sakhelashvili, ``{Electroweak $\eta_w$ meson},'' \href{http://arxiv.org/abs/2408.07535}{{\ttfamily arXiv:2408.07535 [hep-th]}}.

\end{thebibliography}\endgroup

\end{document}